\documentclass[alpha-refs]{wiley-article}

\usepackage{siunitx}

\def\ci{\perp\!\!\!\perp}

\newcommand{\prob}[1]{\mathbb{P}\left(#1\right)}
\newcommand{\ic}[1]{\mathbb{I}\left\{#1\right\}}
\newcommand{\expt}[1]{\mathbb{E}\left(#1\right)}
\newcommand{\n}{\\ \nonumber}
\newcommand{\bx}{\boldsymbol{x}}
\newcommand{\bX}{\boldsymbol{X}}
\newcommand{\Norm}[1]{\left|\left| #1 \right|\right|}

\papertype{Original Article}
\title{Instrumental Variable Methods using Dynamic Interventions}

\author[1\authfn{1}]{Jacqueline A. ~Mauro, PhD}
\author[2\authfn{2}]{Edward H. ~Kennedy, PhD}
\author[3\authfn{3}]{Daniel Nagin, PhD}

\affil[1]{School of Information, UC Berkeley, Berkeley, CA, 94720, USA}
\affil[2]{Department of Statistics, Carnegie Mellon University, Pittsburgh, PA, 15213, USA}
\affil[3]{Heinz College, Carnegie Mellon University, Pittsburgh, PA, 15213, USA}

\corraddress{Jacqueline A. ~Mauro PhD, School of Information, UC Berkeley, Berkeley, CA, 94720, USA}
\corremail{jacqueline.mauro@berkeley.edu}

\presentadd[\authfn{1}]{School of Information, UC Berkeley, Berkeley, CA, 94720, USA}

\runningauthor{Jacqueline A. ~Mauro et al.}

\begin{document}

\maketitle

\begin{abstract}
Recent work on dynamic interventions has greatly expanded the range of causal questions researchers can study. Simultaneously, this work has weakened identifying assumptions, yielding effects that are more practically relevant. Most work in dynamic interventions to date has focused on settings where we directly alter some unconfounded treatment of interest. In policy analysis, decision makers rarely have this level of control over behaviors or access to experimental data. Instead, they are often faced with treatments they can affect only indirectly and whose effects must be learned from observational data. 

In this paper, we propose new estimands and estimators of causal effects based on dynamic interventions with instrumental variables. This method does not rely on parametric models and does not require an experiment. Instead, we estimate the effect of a dynamic intervention on an instrument. This robustness should reassure policy makers that these estimates can be used to effectively inform policy. 

We demonstrate the usefulness of this estimation strategy in a case study examining the effect of visitation on recidivism.

\keywords{Nonparametrics, Causal Inference, Recidivism, Instrumental Variables, Dynamic Interventions}
\end{abstract}

\section{Introduction}\label{sec:Background}
This paper is the only extension, to our knowledge, of dynamic interventions to instrumental variables. We use these developments to study the effects of inmate visitation on future criminal behavior. Beyond helping to answer an important substantive question, our proposed method provides four important contributions: inference is fully nonparametric, the method is valid in observational settings, the method can incorporate continuous instruments, and finally, the method estimates the effect of more realistic interventions than have generally been available, relaxing positivity assumptions and providing practicable decision support. 

Previous evidence from theoretical and empirical research in the criminology literature suggests that visitation reduces recidivism (\citet{Bales2008, Borgman1985, Casey-Acevedo2004, Cochran2014, Derkzen2009, Duwe2013, Holt1972, Mears2012}).  For an excellent review of the literature on recidivism and desistance, see \citet{Bersani2018}. Intuitively, the salutory effect of visitation makes sense: visitation allows prisoners to remain connected to the world outside of prison, and encourages them to feel a sense of responsibility to their family and friends. As is often the case in policy work, however, research into the effects of visitation on recidivism is associational (\citet{Durose2014}), or depends on parametric assumptions (\citet{Bales2008, Cochran2014, Duwe2013}) and/or matching (\citet{Cochran2014, Mears2012}).  These studies are often sensitive to model misspecification or violations of other strong assumptions.

Causal inference in such complex settings must be able to overcome three central challenges. First, experiments are rarely available, and ``no unmeasured confounding'' assumptions may be violated. Thus researchers often resort to using instrumental variables. A second difficulty arises when continuous instruments rather than binary ones are most suitable, since we are rarely interested in studying the effect of setting everyone to a specific instrument level in the non-binary case. Finally, policy settings are complex and often involve many covariates; we have little reason to believe we know a parametric relationship between covariates, instrument, treatment and outcome. The method we propose can satisfy all of these demands, providing robust inference in a complex setting. 

In this paper, we present a case study to illustrate these methods, looking at the effects of visitation on recidivism. In this example, we cannot intervene directly on the treatment--visitation--but we can implement policies that encourage or discourage visits. In addition, we know that visitation is confounded with recidivism rather than assigned randomly or pseudo-randomly, which forecloses many causal methods. Using our estimator, we can robustly and nonparametrically examine the effects of visitation on recidivism among those whose visitation we can affect.

The paper is organized as follows. In the remainder of section \ref{sec:Background}, we describe the relevant statistical and criminological concepts. We give a primer on potential outcomes as well as instrumental variables, and describe common approaches in the causal literature to the parametrization, identification and estimation of causal parameters. In section \ref{sec:Proposed}, we discuss the proposed shift estimator and its properties. We demonstrate these properties through a simulation study in section \ref{sec:Simulation}, and apply the shift estimator to our case study in section \ref{sec:Application}. We conclude with a discussion and future work.

\subsection{Language and Notation of Causal Inference}
Throughout, we will rely on the Neyman-Rubin potential outcomes framework (\citet{rubin1974}). In this setting, one imagines a ``potential outcome'': the outcome we would have seen had we implemented a certain treatment (or after a given exposure). We ask, for example, if a prisoner would have committed a new crime had he been visited and compare this to whether he would have committed a new crime if he were not visited. If there is a difference between these two potential outcomes, we say there is a causal effect of visitation on his recidivism. 

We will also rely on instrumental variables (IVs), which are a fundamental tool in observational causal inference, and are particularly popular in econometrics (\cite{Acemoglu2002,Angrist1996,Heckman1990,Etile2015,Wooldridge1999}). An instrument, denoted by $Z$, is in essence an unconfounded variable which affects the outcome only through the treatment. When the structure of the relationships between the instrument and other variables meet some additional criteria described later, the instrument acts almost as a proxy for randomization and makes causal inference possible without an experiment or in an experiment with noncompliance.

Throughout, we use the following notation. Our observed data are denoted $\mathcal{O} = (Y,A,Z,\bX) \stackrel{iid}{\sim} P$, where $Y$ denotes the outcome of interest (recidivism), $A$ denotes the binary treatment of interest (visitation), $Z$ denotes the instrument (distance in minutes to inmate's family home) and $\bX$ is a set of observed covariates (demographics, crime type, etc.). 

In our approach, a step towards estimating the causal effect is to first estimate so-called ``nuisance parameters.'' We use the term ``nuisance'' because although they are important quantities that feed into our final estimate, we are not directly interested in them. The relevant nuisance quantities are denoted by,
\begin{align*}
    \eta &= \left(\mu(Z,\bX), \lambda(Z,\bX), \pi(Z \mid \bX)\right)
\end{align*}
Where $\mu(z,\bx) = \expt{Y \mid Z=z, \bX = \bx}$ is the outcome regression, $\lambda(z,\bx) = \expt{A \mid Z=z, \bX = \bx }$ is the treatment regression and $\pi(z \mid \bx) = P(Z=z \mid \bX =\bx)$ is the conditional instrument density, (i.e., instrument propensity score function). 

Finally, we follow conventions in causal literature for potential outcomes and denote the potential outcome under both $(A=a, Z=z)$ as $Y^{az}$, similarly $Y^a=Y^{a{Z^a}}$ is the potential outcome under just $A=a$, and $A^z$ is an additional counterfactual introduced by the use of an IV -- the potential treatment under $Z=z$. 

\subsection{Dynamic Interventions}

Causal effects are summaries of outcomes under some intervention on a treatment variable, where every intervention falls in a $2\times2$ classification of types: static vs.\ dynamic, and deterministic vs.\ stochastic. \textit{Static} interventions change treatment in a way that does not depend on any other information (e.g., covariate or past treatment information); \textit{dynamic} interventions are allowed to depend on such other information. \textit{Deterministic} interventions change treatment in a uniform way that is not random; \textit{stochastic} interventions incorporate some randomness. 

Interest in dynamic interventions (both stochastic and deterministic) is growing, because these interventions are often more plausible in practice and because they allow identification under relaxed assumptions compared to deterministic intervention effects like the ATE. For example, \citet{Munoz2012} proposed a particular class of stochastic dynamic interventions for continuous treatments, based on an intervention that replaces the observed treatment with a random draw from the observed conditional treatment distribution (given covariates) shifted by some factor $\delta$. Mathematically, the observed treatment $A$ is replaced with a new $A^*$ drawn from the density $p(A+\delta \mid X)$. Note that this intervention is stochastic since it is a random draw, and it is dynamic because it depends on the observed covariate and treatment. 

\citet{Haneuse2013} recast effects under such interventions as effects of ``modified treatment policies'' and showed that they have a deterministic interpretation, in which the observed treatment is replaced not with a random draw but with the precise value $A^*=A + \delta$, which is completely known given the observed treatment. These deterministic ``modified treatment policies'' ask what would happen if each unique observed treatment value was shifted by the quantity $\delta$. \citet{Haneuse2013} also showed that effects under this deterministic interpretation can be identified under slightly weaker positivity assumptions. \citet{Young2014} studied similar effects of interventions that change treatment depending on how the observed value relates to a relevant threshold. For more examples of dynamic interventions we refer the reader to \citet{Cain2010,Dudik2014,Moore2012,Murphy2003,Pearl2009,Robins2004,Robins2008,Taubman2009,Tian2008,VanDerLaan2007,kennedy2019nonparametric}.

Standard causal inference methods were built for estimating effects of static interventions and require each individual to have some non-zero chance of receiving each level of treatment. This is often called a ``positivity'' or ``experimental treatment assignment'' assumption. In reality, there are often some subjects who would never receive certain interventions; in the example we discuss shortly, for example, it would never be possible to ensure that every prison inmate was located exactly 20 minutes away from his or her family home. Some family homes are more than 20 minutes from every prison. Dynamic interventions can relax the strong positivity assumptions required to identify causal effects; for example in the intervention proposed by \citet{Haneuse2013}, one only needs each subject to have some chance at receiving the treatment level $\delta$ units away from their observed value, instead of more extreme levels that are further away from what was actually observed.

 The aforementioned work on dynamic effects has expanded the field of causal inference to incorporate more feasible interventions. However, previous work has relied exclusively on a strong exchangeability or no unmeasured confounding assumption, i.e., that treatment is randomized given observed covariates. This may be reasonable in some (e.g., medical) settings, but is often implausible in policy or other settings. In these cases, the assumptions required for an instrumental variables strategy like the one proposed here may be more plausible, since they do not require an unconfounded treatment. In our case, prison administrators have considerable control over the distance prisoners will be from their next of kin. This can therefore be modified by a policy change and is more likely to be explained by measured variables, i.e., less likely to be confounded. In contrast, prison administrators have no direct control over the complex process determining whether and how prisoners receive visits, except in their ability to forbid it. 
 
The primary goal of our paper is to fill the gap between policies of interest and the available statistical methods by incorporating dynamic interventions with IVs. This approach yields more relevant effects under weaker identifying assumptions and allows the researcher to use nonparametric models without sacrificing fast convergence rates.

\section{Proposed Approach}\label{sec:Proposed}
\subsection{Parametrization of the Estimand}
The first step in the analysis of a causal effect is to decide what causal effect we actually want to learn. The Local Average Treatment Effect (LATE) with a binary $Z$ is a common example. In this case, we want to learn the effect of treatment $A$ on individuals who receive treatment if assigned $Z=1$ but do not if $Z=0$. When $Z$ is not binary, researchers might seek to estimate a version of the LATE among compliers who take treatment at some specific $Z=z$ but not at another $Z=z'$,
\begin{align}
    \label{eq:LATEbad}
    \psi_{LATE} &= \mathbb{E}\{ Y^{1} - Y^{0} |  A^{z}>A^{z'} \}
\end{align}

This parameter presents three challenges not posed by the usual LATE parameter: 1) substantively, it does not relate to an intuitive intervention--could one actually set $Z=z$ or $Z=z'$ for all subjects?; 2) it is difficult to estimate because it is not pathwise differentiable and does not admit $\sqrt{n}$-consistent estimation nonparametrically; and 3) it requires stronger identifying assumptions than the usual LATE, because of the continuous IV.

Another approach to dealing with non-binary $Z$ is to take the classical IV approach, often a two-stage least squares or some variant, which estimates the LATE as a parameter in a regression. This model posits,

\begin{align}
    \label{eq:TSLS}
    \expt{A \mid Z, \bX} &= \alpha_0 + \alpha_1 \bX + \alpha_2 Z \\
    \expt{Y \mid A, \bX} &= \beta_0 + \beta_1 \bX + \beta_2 A\nonumber
\end{align}

This parametric approach is popular, but relies on the strong assumption that the researcher can explicitly describe the relationships between possibly high-dimensional covariates, instrument, treatment and outcome. We will show that the use of influence-function based estimation allows us to forgo these assumptions and still estimate the effect of any number of dynamic interventions. 

In order to define the causal effect of interest, we need to think about the intervention under study. In the ``Usual LATE'' case given in (\ref{eq:LATEbad}), the intervention is to set $Z=z$ for everyone  and compare that to setting $Z=z'$ for everyone. More generally, let us denote interventions by $h$, functions of $Z$ which may depend on $\bX$ and may be stochastic or deterministic. Some examples of such functions are:
\begin{align}
    \text{"Usual LATE": } h_1(Z;\bX) = z &; \ \ h_2(Z;\bX) = z' \\
    \text{"Single Shift": } h_1(Z;\bX) = Z + \delta &; \ \ h_2(Z; \bX) = Z; \ \ \delta \in \mathbb{R} \n
    \text{"Double Shift": } h_1(Z;\bX) = Z + \delta &; \ \ h_2(Z; \bX) = Z - \delta; \ \ \delta \in \mathbb{R} \\ \label{hDoubleShift}
    \text{"Random Shift": } h_1(Z;\bX) = Z + \delta_1(\bX) &;h_2(Z; \bX) = Z + \delta_2(\bX)\\
    \delta_1 \sim Unif(0,|X|)&,\ \ \delta_2 \sim Unif(-|X|,0)\\
    \text{``Modified Treatment Policy'': } h_1(Z;\bX) &= Z\ic{Z\geq 30} + 30\ic{Z \leq 30}\n
    h_2(Z;\bX) &= Z 
\end{align}

The effect of interest is then generally defined as,
\begin{align*}
    \psi_h = \expt{Y^1 - Y^0 \mid A^{h_1(Z; \bX)}-A^{h_2(Z; \bX)}}
\end{align*}

Each intervention defines a complier population and will lead to estimates of treatment within that population. The single shift intervention, for example, defines a complier population as those who receive treatment when their instrument increases by a set amount $\delta$ but do not at its current level. The double shift complier population are those who receive treatment when their instrument is increased by $\delta$ but do not when the instrument is decreased by $\delta$. Any compliers in the single shift are therefore also included in the double shift complier population for a given $\delta$. 

The choice of $h$ functions depends on the intervention of interest. If we want to imagine a policy that encourages administrators to prioritize keeping prisoners close to home, we may want to use the double shift estimand (\ref{hDoubleShift}), which shifts the entire distribution of $Z$ (distance from home) by some amount $\delta$ in each direction from where we observe it currently. 

In this paper, all results will be given for the treatment effect among compliers with respect to the double shift intervention above, i.e., 
\begin{align}\label{eq:DoubleShift}
    \psi_{shift} &= \expt{Y^1 - Y^0 \mid A^{Z + \delta} > A^{Z-\delta}}
\end{align}
when $Z$ has unbounded support on the whole real line; to the best of our knowledge, this kind of IV causal effect has not yet been proposed in the literature. In the case where the support of $Z$ is a bounded interval $[z_{min},z_{max}]$, we consider a slightly modified version
\begin{align}\label{eq:DoubleShift2}
    \psi_{shift}^* &= \expt{Y^1 - Y^0 \mid A^{Z + \delta_u^*} > A^{Z-\delta_\ell^*}}
\end{align}
for $\delta_u^*=\delta \ic{Z \leq z_{max}-\delta}$ and $\delta_\ell^*=\delta \ic{Z > z_{min}+\delta}$. The latter estimand simply considers compliers whose observed $Z$ values are within $\pm \delta$ of the bounds on the support, i.e., those subjects for whom a shift intervention on the IV is feasible.

Conceptually, these effects can be thought of in one of two ways. In the first, we imagine an intervention which takes each individual's observed distance from their next of kin $Z$ and increases or decreases it by an amount $\delta$. We then estimate the effect of this intervention on the recidivism of those whose visitation changes when the distance is changed. In the second conception, the intervention shifts the current distribution of distances and draws new values of $Z$ from that shifted distribution. We then examine the effect of visitation among those whose visitation changes between the original and shifted distributions of distance. 

The second interpretation may be more attractive in a policy setting, because it roughly corresponds to a policy change that would direct administrators to prioritize keeping inmates closer to home or facilitating transportation. 

It is relatively straightforward to move from the shift case to the more general $h$ case, although some subtleties arise, especially around positivity and the conditions that must be imposed on the $h$ functions. See \citet{Haneuse2013} for an idea of what the general case will look like.

\subsection{Identification}
Having chosen a causal parameter to estimate, the fundamental problem of causal inference arises: the goal is to make inference about a functional of the distribution of potential outcomes, but these counterfactuals are only partially observed. The procedure to move from a functional of unknown potential outcome distributions to an estimable functional of the observational distribution $P$ is called identification. 

 We will use the following base assumptions for identifying versions of the IV estimands mentioned in the previous subsection:
\begin{enumerate}
    \item Consistency: $Z = z \Rightarrow A = A^z$ and $(Z,A) = (z,a) \Rightarrow Y = Y^{za}$
    \item Ignorability of $Z$: $Z \ci (Y^{z}, A^{z})\mid X$
    \item Exclusion Restriction: $Y^{za} = Y^a$
    \item Monotonicity: $\prob{A^{z} \ge A^{z'}} = 1$ for $z>z'$
\end{enumerate}

We will also require instrumentation and positivity assumptions, which we will describe shortly, and which depend on the particular form of the double shift parameter. Assumptions 1--4 are common to the IV literature. It is important to note that having made these assumptions and identified our parameter in the data, we have moved from a causal question to a purely statistical one. The positivity assumption required is determined by the effect under study. 

Note that the ignorability assumption is not saying that $Z$ is independent of $A$ given $X$. Clearly, that would create problems for our entire procedure because it would require that the instrument have no effect on the treatment given covariates. Instead, we assume $Z$ is conditionally independent of the potential outcomes of $A$, which will hold if $Z$ is as good as randomized within covariate strata. It would be violated if, for example, inmates are moved closer to home for good behavior, and good behavior was related to the probability of recidivating. 

Under the assumptions 1--4, the LATE parameter with a binary $Z$ can be identified as:
\begin{align}
    \label{eq:LATEid}
    \psi_{LATE} &= \expt{Y^1 - Y^0 \mid A^1 > A^0} = \frac{\expt{\mu(1,\bX) - \mu(0, \bX)}}{\expt{\lambda(1,\bX) - \lambda(0, \bX)}}
\end{align}

The positivity assumption needed in this case is:
\begin{align}
    \label{eq:UsualPos}
    \mathbb{P}(\epsilon < \pi(z \mid \bX) < 1-\epsilon) = 1
\end{align}
This positivity assumption requires that each subject have positive probability of receiving all levels of $z$. In the binary $Z$ case, this means both $Z=0$ and $Z=1$ must have positive probability. When $Z$ is continuous or multivalued rather than binary, this assumption becomes stronger, because it says each subject must have positive probability of receiving each value of $z$. We will refer to (\ref{eq:UsualPos}) as the ``usual'' positivity requirement and we will show that the estimator we propose relaxes this assumption in important ways.

For the case of the double shift interventions on $Z$, we make the instrumentation assumption,
\begin{equation} \label{eq:DSinstrumentation}
P(A^{Z+\delta} > A^{Z-\delta})>0 
\end{equation} 
and the positivity assumption,
\begin{align}\label{eq:DSPositivity}
    \prob{\epsilon < \pi(Z + \delta \mid \bX) < 1-\epsilon} = 1\\
    and \ \ \prob{\epsilon < \pi(Z - \delta \mid \bX) < 1-\epsilon} = 1 \nonumber
\end{align}

This positivity assumption implies that if we observe $Z=z$, there must be a non-zero chance of observing both $z \pm \delta$ (replacing $\delta$ with $\delta^*_u, \delta^*_l$ as needed). Note that this is weaker than the assumption in \ref{eq:UsualPos}, because the instrument is not required to take on every value in its support, only values around its observed value. The indicator function ensures that the positivity requirement holds when $z_{max}, z_{min}$ are known, but for convenience, we may drop the indicator function and assume the support of $Z$ is unbounded. For the proof of this (and all other proofs) see the Appendix.

Under Assumptions 1--4, instrumentation (\ref{eq:DSinstrumentation}), and positivity (\ref{eq:DSPositivity}) the double shift LATE parameters are identified as,
\begin{align} \label{eq:singleID}
\psi_{shift} &= \frac{\expt{\mu(Z+\delta,\bX) - \mu(Z-\delta, \bX)}}{\expt{\lambda(Z+\delta,\bX) - \lambda(Z-\delta, \bX)}}
\end{align}
\begin{align} \label{eq:singleID2}
\psi_{shift}^* &= \frac{\expt{\mu(Z+\delta^*_u,\bX) - \mu(Z-\delta^*_l, \bX)}}{\expt{\lambda(Z+\delta^*_u,\bX) - \lambda(Z-\delta^*_l, \bX)}}
\end{align}
This is similar to the usual LATE, except that instead of getting predicted values under $Z=0$ and $Z=1$ we do so under shifted versions of the observed $Z$.

As noted earlier, the positivity assumptions required for identifying $\psi_{shift}$ are substantially weaker than those required for typical LATEs. For every set of covariates $\bX$ with positive probability, the usual positivity assumption says we must be able to observe any value $z \in Supp(Z)$. This would imply, for example, every prisoner must have positive probability of being at \emph{every} distance from home within the support of $Z$. 

In contrast, identifying the shift estimand requires an assumption about shifts around the observed values of $Z$, instead of the likelihood we could attain an arbitrary $z$ for all subjects. So each subject need not have positive probability across a wide range of values $z$, but only for their own observed value (which is automatic), and for a value that is shifted away from their observed value.

We can illustrate this with a simple simulation. Imagine our data-generating process is a mixture of truncated normals, with means dependent on a single binary covariate $X$. 
\begin{align}
    X &\sim Bern\left(1/2 \right)\\
    Z &\sim TruncNorm\left(\mu = 2X-1, \sigma = 0.5, a = -X - 3(1 - X), b = 3X + 1(1 - X) \right)
\end{align}

In our example, we could imagine that rural ($X=0$) inmates are on average slightly farther from home than urban ($X=1$) inmates. Distance is continuous in each case, but there are hard maximum and minimum values for each population at the nearest and farthest prisons. It may be that we do not know these edges or how they depend on covariates, and therefore cannot set up our estimates to account for this. Although in this setting we have focused on $\psi_{shift}$ these results hold for $\psi^*_{shift}$ by the same logic as presented here.

In the usual positivity setting, we require all $z$ within the support $[-3,3]$ to have positive probability. However, we know that for any subject with covariates $X=1$, all $z<-1$ have zero probability. Likewise, for all subjects with $X=0$, any value of $z>1$ has zero probability. 

In the shift case, however, the positivity assumption will only be violated if we draw observations where a shift up or down pushes $Z$ outside of its support conditional on $X$. For example, if $X=0$ and we draw a $Z$ value such that $Z < -3 + \delta$, we know that $Z-\delta < -3$ will have zero probability and will be a positivity violation. If we set $\delta = .1$, for example, the chances of drawing such an observation are extremely small. 

We illustrate this in Figure (1). The light boxes outline positivity violations in the usual setting. The narrow dark boxes represent an area where observations would violate positivity. In our simulations ($n=5000, \delta = 0.1$), there were no violations of the shift parameter's positivity assumption. 

\begin{figure}[h!]
	\begin{center}
	\includegraphics[width=.75\textwidth]{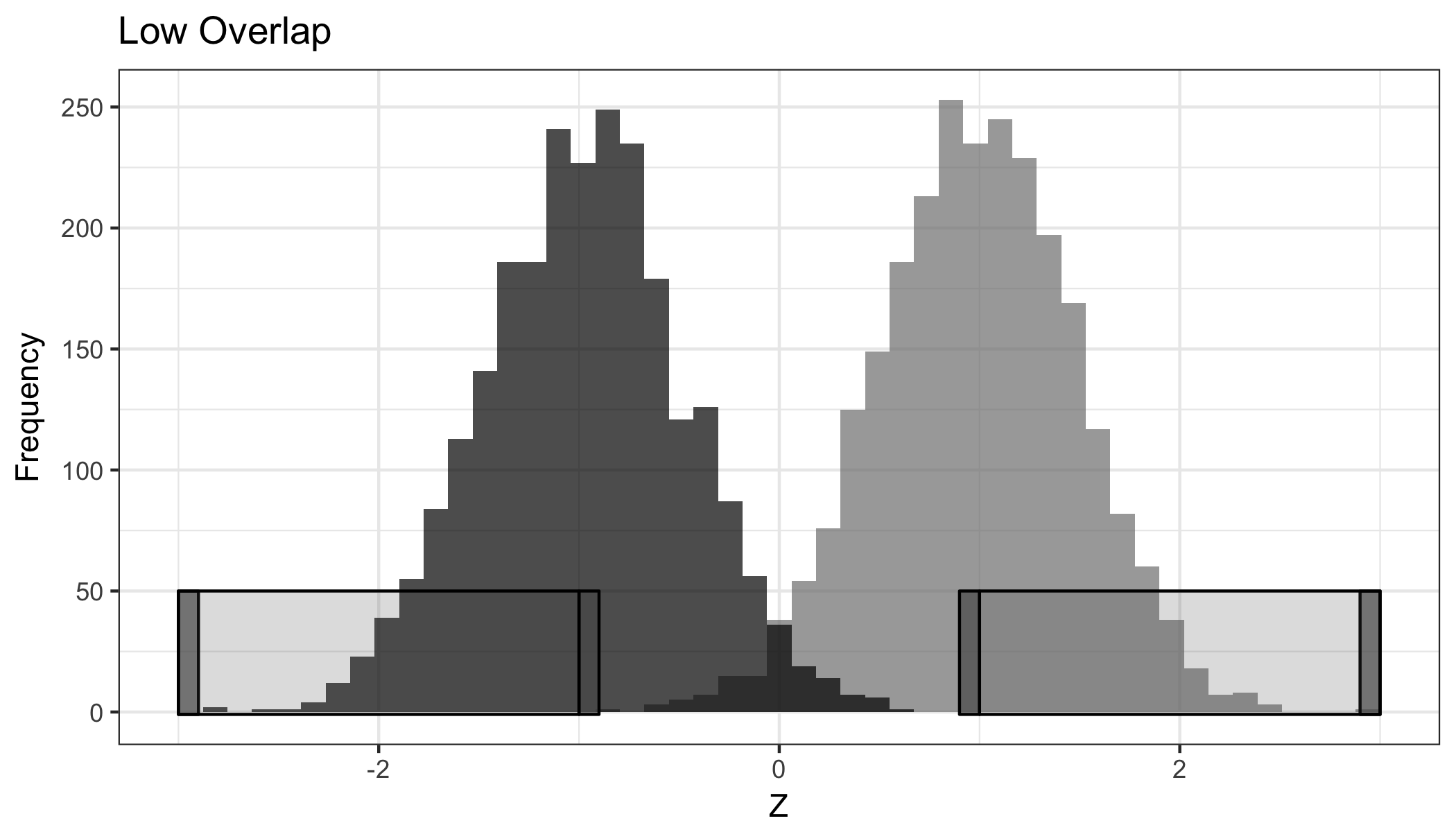}
	\caption{\small Values of $Z$ or $z$ which violate positivity. Dark narrow rectangles show the range of observed draws of $Z=z$ that would violate our positivity assumption. Lighter long rectangles show the range of $Z$ that is a violation of usual positivity requirements.}
	\end{center}
	\label{fig:posEx}
\end{figure}

\subsection{Estimation}
The next step is to estimate the identified parameter (in our case, \eqref{eq:singleID} or \eqref{eq:singleID2}). This is often accomplished in one of three ways: through regression based estimators, weighting-based estimators or doubly-robust estimators. Estimation has historically relied upon parametric models like linear or logistic regression (\cite{Wooldridge1999}). The two-stage least squares approach in (\ref{eq:TSLS}) is one example of this approach. In the parametric setting, the relationship between covariates, instrument, treatment and outcome is assumed known (for example, it is assumed to be linear). This assumption may be too strong in many cases, especially in contexts with a large number of covariates or a complex system.

Regression-based estimators take parameters like that in equation (\ref{eq:singleID}) and simply plug in estimates of the regression functions. For this reason, we sometimes refer to them as ``plug-in'' estimators. The regression-based estimator of the shift parameter is therefore:
\begin{align}
    \hat{\psi}_{reg} = \frac{\hat{\mathbb{E}}\left\{ \hat{\mu}(Z+\delta,\bX) -  \hat{\mu}(Z-\delta,\bX)\right\} }{\hat{\mathbb{E}}\left\{ \hat{\lambda}(Z+\delta,\bX) -  \hat{\lambda}(Z-\delta,\bX)\right\} }
\end{align}

In order to estimate this, one might assume that the regression function $\mu(Z+\delta,\bX) = \expt{Y \mid Z+\delta,\bX}$ is a linear function of the covariates. $\hat{\mu}(Z+\delta,\bX)$ is then predicted by taking a linear regression of $Y$ on the covariates. Averaging the predictions from these regression functions give the estimate of $\psi_{shift}$. Alternatively one could use nonparametric estimators of these regression functions, e.g., based on kernel smoothing, random forests, etc.  

A weighting-based approach is similar, but instead of modelling the conditional means of the outcome and treatment variables, the propensity scores -- the chances of receiving treatment conditional on covariates-- are modelled. This yields estimators of the form:
\begin{align}
    \hat{\psi}_{IPW} &=\frac{\hat{\mathbb{E}}\left\{  \left(\frac{\hat{\pi}(Z-\delta \mid \bX)-\hat{\pi}(Z+\delta \mid \bX)}{\hat{\pi}(Z \mid \bX)}\right)Y \right\}}{\hat{\mathbb{E}}\left\{  \left(\frac{\hat{\pi}(Z-\delta \mid \bX)-\hat{\pi}(Z+\delta \mid \bX)}{\hat{\pi}(Z \mid \bX)}\right)A \right\}} 
\end{align}

These estimators have good properties when the propensity scores (the $\pi$ terms) are known, as in an experiment. When they must be estimated, however, these models become vulnerable to model misspecification and if estimated nonparametrically, will not attain $\sqrt{n}$-rates except under specific circumstances (e.g., for particular and undersmoothed nuisance estimators).

Modern methods have shed some light on how to overcome some of these concerns. Notably, influence function (IF) based estimation frees researchers from many of the untenable parametric assumptions of earlier methods. We say an estimator has a particular influence function when the estimator (minus its target) is equivalent to a sample average of the influence function, up to $o_P(1/\sqrt{n})$ error. A somewhat deeper and more nuanced notion of influence function concerns the influence function of a \emph{parameter}; a parameter has a particular influence function if the parameter admits a von Mises or distributional Taylor expansion, with the influence function acting as the derivative term in the expansion. For more details on influence functions, see \citet{Tsiatis2006, Bickel1975, VanderLaan2003, vanderVaart1998AsymptoticStatistics, kennedy2017semiparametric}  or Cuellar and Mauro, 2018 (in progress). 

In practice IF-based estimators will often be of a form that blends the regression and weighting approaches. Our next result gives the influence functions for the two double-shift LATE parameters we consider, which motivates appropriate estimators that can attain fast convergence rates even when the nuisance estimators converge at slower nonparametric rates. The variance of the efficient influence function also acts as a benchmark for the smallest variance of any regular asymptotically linear estimator, as well as a local minimax lower bound on the mean squared error \citep{Bickel1975}.

\begin{theorem}\label{thm:doubleIF}
Under the causal conditions and the positivity condition given in (\ref{eq:DSPositivity}), the efficient influence function of the functional $\psi_{shift}^*$ defined in (\ref{eq:DoubleShift2}) is given by

\begin{align}\label{eq:DoubleShiftEff}
\varphi(z; \eta, \psi) &= \frac{ \Xi(Y; \delta^*_u, \delta^*_l )  - \psi  \Xi(A; \delta^*_u, -\delta^*_l )}{\expt{\lambda(Z+\delta^*_u,
\bX) - \lambda(Z-\delta^*_l,\bX)}} 
\end{align}
where $\Xi(T; a, b) \equiv \xi(T;a) - \xi(T; b)$ for $\xi(T; \delta) = \xi_{\mathbb{P}}(T; \bX,Z,\delta) \equiv \frac{\pi(Z-\delta|\bX)}{\pi(Z|\bX)} \{T - \expt{T|\bX,Z}\} + \expt{T|\bX,Z + \delta} $ the (uncentered) efficient influence function for the mean of arbitrary $T$ (here replace $T$ with either $A$ or $Y$) under a stochastic/dynamic intervention that sets $Z$ to $Z+\delta$.  

Recall $\delta^*_u = \delta \ic{Z \leq z_{max} - \delta}$, $\delta^*_l = \delta \ic{Z \geq z_{min} + \delta}$. When the support of $Z$ is unbounded, the efficient influence function for $\psi_{shift}$ is the same as above except with $\delta^*_u, \delta^*_l$ replaced with simply $\delta$.
\end{theorem}
For the proof of this theorem, see the Appendix section \ref{sec:EIF}. 

A standard approach for constructing efficient estimators based on influence functions is to solve an estimating equation using the estimated influence function as an estimating function. That is, we solve for $\hat{\psi}$ in, 
\begin{align*}
    \mathbb{P}_n (\varphi(z; \hat{\eta}, \hat{\psi})) = 0
\end{align*}

Based on the formulation of $\varphi$ given in the above theorem, we propose the IF-based estimator:
\begin{align}
\label{DoubleEstimation}
\hat{\psi^*}_{IF} &= \frac{ \mathbb{P}_n \left[\ \widehat\Xi(Y; \delta^*_u, -\delta^*_l ) \right] } { \mathbb{P}_n \left[  \widehat\Xi(A; \delta^*_u, -\delta^*_l )\right]}
\end{align}
for $\widehat\Xi$ an estimate of $\Xi$. Similarly, for our double shift parameter with unbounded support, the corresponding IF-based estimator is given by,
\begin{align}
\label{GeneralEst}
\hat{\psi}_{IF} &= \frac{ \mathbb{P}_n \left[\ \widehat\Xi(Y; \delta, -\delta ) \right] } { \mathbb{P}_n \left[  \widehat\Xi(A; \delta, -\delta)\right]}
\end{align}

The remaining task is simply to pick a nonparametric method (or set of methods) to estimate the nuisance parameters. We will show in the following section that nonparametric rates on the nuisance parameters are sufficient to achieve parametric rates on the causal estimand overall. In practice, we use SuperLearner, an ensemble learner, to estimate the $\mu$ and $\lambda$ parameters as well as possible. To estimate the conditional density, we rely on random forests and a Gaussian kernel. We chose these tools for maximum flexibility, but the user is relatively free to pick different tools if they prefer. Appendix section \ref{sec:Estimation} gives more detail into the strategy we used and the conditions that need to be met in nuisance parameter estimation. 

Note that another increasingly popular approach to nonparametric causal inference involves the use of Targeted Maximum Likelihood Estimation (TMLE). This method similarly uses machine learning to estimate nuisance parameters and relies on the influence function. It differs from estimating equation approaches by using an additional targeting step to create a plug-in estimator, ensuring that the estimator respects the bounds of the parameter space. The ratio structure of our parameter likely makes such a TMLE slightly more complicated than that for an unconfounded average treatment effect; developing such an approach would be valuable future work.

In the next subsection we study the large-sample properties of our proposed estimators, and show that they can attain fast $\sqrt{n}$-rates of convergence, even when built from flexible nonparametric machine learning tools.

\subsection{Asymptotic Properties}
In this section we highlight some of the notable properties of our proposed IF-based estimator of the shift parameter $\psi_{shift}$, namely that they are doubly robust and can converge at faster rates than the nuisance estimators upon which they rely. This makes them less sensitive to the curse of dimensionality than simple plug-in IPW/regression estimators. Although we focus on estimating $\psi_{shift}$, the same properties also hold for the LATE $\psi_{shift}^*$ based on feasible IV interventions. Throughout, we let $\Norm{ f } = \int f(z)^2 \ dP(z)$ denote the squared $L_2(P)$ norm.

\begin{theorem}\label{thm:DoubleDR}
Dropping the distinction between the bounded and unbounded cases for ease of notation, suppose the following conditions hold:
\begin{enumerate}
    \item The nuisance function and their estimators belong to a Donsker class.
    \item Strengthened Positivity holds: $\prob{\epsilon < \frac{\pi(Z \pm \delta \mid \bX) }{\pi(Z\mid \bX)} < C}=1$ for some $\epsilon >0 $ and $C<\infty$.
\end{enumerate}
Then, 
\begin{align}
    \begin{split}
    \hat{\psi}_{IF} - \psi_{shift} &= O_p\left[\frac{1}{\sqrt{n}}+ \Norm{\frac{\hat{\pi}(Z-\delta \mid \bX)}{\hat{\pi}(Z\mid \bX)} - \frac{\pi(Z-\delta \mid \bX)}{\pi(Z \mid \bX)}} \right. \\ 
    &\cdot \left.  \left(  \Norm{\mu(Z+\delta ,\bX) - \hat{\mu}(Z+\delta , \bX)} +\Norm{\lambda(Z+\delta ,\bX) - \hat{\lambda}(Z+\delta ,\bX)}  \right) \right. \\
    &+ \left. \Norm{\frac{\hat{\pi}(Z+\delta \mid \bX)}{\hat{\pi}(Z\mid \bX)} - \frac{\pi(Z+\delta \mid \bX)}{\pi(Z \mid \bX)}} \right. \\ 
    &\cdot \left.  \left(  \Norm{\mu(Z-\delta ,\bX) - \hat{\mu}(Z-\delta , \bX)} +\Norm{\lambda(Z-\delta ,\bX) - \hat{\lambda}(Z-\delta ,\bX)}  \right) 
    \right]
    \end{split}
\end{align}

If further
\begin{align}
    \Norm{\frac{\hat{\pi}(Z-\delta \mid \bX)}{\hat{\pi}(Z \mid \bX)} - \frac{\pi(Z-\delta \mid \bX)}{\pi(Z \mid \bX)}} &\cdot \left(  \Norm{\mu(Z+\delta, \bX) - \hat{\mu}(Z+\delta, \bX)} +\Norm{\lambda(Z+\delta, \bX) - \hat{\lambda}(Z+\delta, \bX)}  \right) \\
    + \Norm{\frac{\hat{\pi}(Z+\delta \mid \bX)}{\hat{\pi}(Z \mid \bX)} - \frac{\pi(Z+\delta \mid \bX)}{\pi(Z \mid \bX)}}&\cdot \left(  \Norm{\mu(Z-\delta, \bX) - \hat{\mu}(Z-\delta, \bX)} +\Norm{\lambda(Z-\delta, \bX) - \hat{\lambda}(Z-\delta, \bX)} \right) \\
    &=o_p(1/\sqrt{n})
\end{align}
then we have,
\begin{align}
    \label{eq:ConvToNormal}
    \sqrt{n}(\hat{\psi}_{IF} - \psi_{shift}) \rightsquigarrow N(0,\sigma^2)
\end{align}
\end{theorem}

This shows that the asymptotic variance of $\psi_{shift}$ is given by the variance of $\varphi$ and is therefore easy to estimate, allowing simple closed-form asymptotically valid confidence intervals. Specifically we estimate the asymptotic variance simply by computing the empirical variance of the estimated influence function. 

In addition, the estimator $\hat{\psi}_{IF}$ will converge at a $\sqrt{n}$ rate and will be optimally efficient if the product of the nuisance error rates converges to zero faster than $n^{1/2}$, for example if they each converge at faster than $n^{1/4}$. This $n^{1/4}$ rate requirement does not require parametric model assumptions, and in fact can hold under general smoothness, structural (e.g., generalized additive model), or sparsity assumptions, allowing a wide variety of modern machine learning tools to be used, while still providing classical inferential guarantees. 

Finally, this theorem shows that the estimator $\hat{\psi}_{IF}$ is doubly-robust, since if either the $\hat{\pi}$ estimator or if the regression estimators $(\hat\lambda,\hat\mu)$ are estimated consistently (not necessarily both), then the estimator $\hat\psi_{IF}$ is consistent. Importantly, if we estimate the nuisance functions $(\pi,\lambda,\mu)$ on a separate independent sample, the Donsker assumption can be avoided entirely, and one only needs consistency of the nuisance functions at any rate for (\ref{eq:ConvToNormal}) to hold.

Although the above theorem is useful for pointwise inference at particular $\delta$ values, in many cases one may prefer confidence bands across a continuous range of shifts. That is, we want a 95\% confidence band that holds across the full range of $\delta$'s under investigation simultaneously, rather than for a single $\delta$. These sorts of confidence bands are termed ``uniform,'' because they cover the true parameter uniformly across a range of measurements. For such function-valued parameters, the following theorem provides uniform confidence bands:

\begin{theorem}\label{thm:Uniform}
Assume the conditions in Theorem \ref{thm:DoubleDR} and that
\begin{enumerate}
    \item $\varphi$ is Lipschitz in $\delta$
    \item $\Norm{\hat{\sigma}(\delta) / \sigma(\delta)}_{\mathcal{D}} -1 = o_p(1)$
    \item  $\sup_{\delta \in \mathcal{D}}  \Norm{\frac{\hat{\pi}(Z-\delta \mid X)}{\hat{\pi}(Z\mid X)} - \frac{\pi(Z-\delta\mid X)}{\pi(Z\mid X)}} \cdot \left(  \Norm{\mu(Z+\delta, X) - \hat{\mu}(Z+\delta, X)} +\Norm{\lambda(Z+\delta,X) - \hat{\lambda}(Z+\delta,X)}  \right) =o_p(1/\sqrt{n})$
    \item  $\sup_{\delta \in \mathcal{D}}  \Norm{\frac{\hat{\pi}(Z+\delta \mid X)}{\hat{\pi}(Z\mid X)} - \frac{\pi(Z+\delta\mid X)}{\pi(Z\mid X)}} \cdot \left(  \Norm{\mu(Z-\delta, X) - \hat{\mu}(Z-\delta, X)} +\Norm{\lambda(Z-\delta,X) - \hat{\lambda}(Z-\delta,X)}  \right) =o_p(1/\sqrt{n})$
\end{enumerate}

Then,
\begin{align}
\sup_{\delta \in \mathcal{D}} \Bigm| \frac{\sqrt{n}(\hat{\psi}_{IF}(\delta) - \psi_{shift}(\delta))}{\hat{\sigma}(\delta)} - \sqrt{n}(\mathbb{P}_n - \mathbb{P})\left( \frac{\varphi(z; \eta, \delta)}{\sigma(\delta)}  \right) \Bigm| = o_p(1/\sqrt{n})
\end{align}
\end{theorem}
This result shows that uniform confidence bands for our estimators across a continuous range of $\delta$ values are possible under assumptions not much stronger than the pointwise requirements. In particular, under these assumptions, the multiplier bootstrap can be used to construct uniform confidence bands \citep{kennedy2019nonparametric}.

\section{Simulations}\label{sec:Simulation}
We run the following simulation, similar to that used by \citet{kennedy2019nonparametric}. Note that for these estimators, however, it is preferred that we have an unbounded support for $Z$, although as we have seen, setting a $z_{max}$ and $z_{min}$ value within the functional allows us to easily adapt to the bounded support setting.
\begin{align}
\label{eq:simulation}
(Y^0,\bX)&\sim N(0, \boldsymbol{I}_5)\\ \nonumber
Z|\bX,Y^0 &\sim N(\alpha^T \bX,2) \\ \nonumber
A &= \mathbb{I}\{ Z \ge Y^0 \} \\ \nonumber
Y &= Y^0 + \psi A
\end{align}

For these simulations, we set $\alpha = (1,1,-1,-1)$ and our causal effect $\psi = 2$. This obeys all the identifying assumptions we have made, and does not involve particularly complex models or high dimensions. 

To examine convergence rates, we use the true forms of our nuisance parameters and add noise of the form $\frac{\epsilon(Z)}{N^{1/k}}$ where we add error of the form $\epsilon(z) \sim N(z,1)$ to the $\mu(z), \lambda(z)$ terms\footnote{This error form may not be obvious compared to something like $\epsilon \sim N(1,1)$. This latter type of additive error will cancel out in the double shift estimator and both the plug-in and IF-based estimators will perform as if there were no error. On the other hand, taking asymptotic error like $\mu(Z)-\Hat{\mu}(Z) \sim N(1,1)$ and $\mu(Z+\delta) - \Hat{\mu}(Z+\delta) \sim N(2,1)$ is incoherent across all $Z$ values.} and $\epsilon \sim N(1,1)$ to the $\pi$ terms. $k$ is the rate at which we estimate our parameters. So when $k=2$, for example, we have $\hat{\mu} = \mu + \frac{\epsilon}{N^{1/2}}$, meaning we are estimating $\mu$ at roughly a $\sqrt{n}$ rate. 

We compare IF-based estimators and a plug-in estimator using these perturbed nuisance parameters. The plug-in estimator is roughly equivalent to a two stage least squares type estimator, adapted to estimate the same parameter as the shift estimator. This type of simulation has two fundamental advantages: 1) it saves a great deal of computation and more importantly, 2) it allows us to exactly control the error rates on the nuisance parameters. Were we to run a simulation based on existing TSLS and IF-based estimation algorithms, we would not know for sure the error rates on the nuisance parameters. 

We run 500 simulations with a sample size of 100, 1,000, 5,000 and 10,000 for $\delta \in [0.5:4]$ and $k \in \{2,3,4,6\}$. The results at 5,000 are shown in Figure \ref{fig:estComp}, the others are given in the appendix. As theory would dictate, both the plug-in and the IF-based estimators do well when our regression and propensity scores are estimated at $n^{1/2}$. However, we see that when the rate is slower than that, the plug-in starts to experience major bias. The IF-based estimator, on the other hand, maintains its performance until the rate is slower than $n^{1/4}$. 

\begin{figure}[h!]
	\begin{center}
	\includegraphics[width=\textwidth]{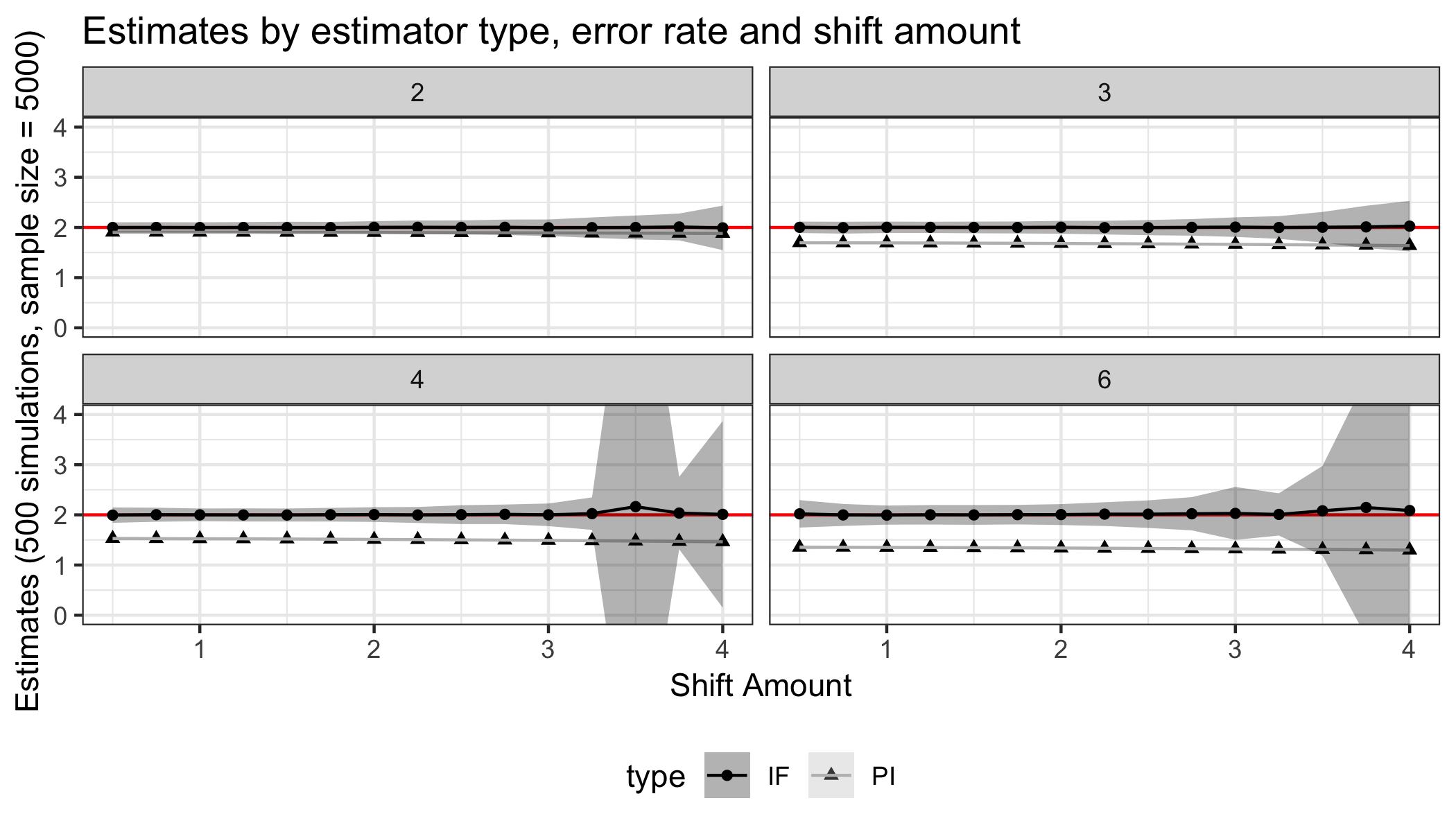}
	\caption{Plug in and Influence-Function Based Estimators across error rates and $\delta$ values}
	\label{fig:estComp}
	\end{center}
\end{figure}

In this figure, we show the empirical standard deviation over the simulations. In practice, we can construct confidence bands using the multiplier bootstrap. 

\section{Application}\label{sec:Application}
We now apply the double shift estimator to the question of whether visitation can affect recidivism risk. This intervention imagines shifting prisoners both nearer and farther from their home, and estimates to what degree the change in their visitation due to this shift changes the chances of their being arrested for a new crime within three years of release. Summary statistics are given in Table \ref{table:visitsDemogs}.  In general, prisoners who are visited are slightly less likely to recidivate (45\% compared to 50\%). They are also nearer to home on average (155 minutes compared to 195 minutes). However, they score more favorably on a number of other covariates which are likely associated with reduced recidivism -- marital status, number of prior incarcerations, etc. -- suggesting there are unmeasured confounders contributing to the observed association between visitation and recidivism. This reinforces the need for a robust causal estimation strategy.

\begin{table}[ht]
\caption{Demographics by Visitation}
\label{table:visitsDemogs}
\centering
\begin{tabular}{lrr}
  \hline
 & Never Visited & Visited \\ 
\hline
Distance from Home (minutes) & 195.11 & 155.17 \\ 
Recidivate & 0.50 & 0.45 \\ 
White & 0.37 & 0.48 \\ 
Days at Final Prison & 360.50 & 452.89 \\ 
Urban & 0.79 & 0.75 \\ 
Number of Prior Arrests & 9.04 & 6.90 \\ 
Number of Prior Incarcerations & 0.55 & 0.31 \\ 
Married & 0.11 & 0.15 \\ 
Violent & 0.22 & 0.28 \\ 
LSIR Score & 25.53 & 22.81 \\ 
Number of Visits & 0.00 & 14.32 \\ 
Custody Level & 2.62 & 2.44 \\ 
On Mental Health List & 0.17 & 0.13 \\ 
High School Grad & 0.56 & 0.63 \\ 
\hline
\end{tabular}
\end{table}

The dataset has been supplied and scrubbed by the Pennsylvania Department of Corrections (DOC). The data include information on all male prisoners released from the Pennsylvania department of corrections in 2008. It contains a unique identifier for each of 5,749 prisoners, the facility in which they were last imprisoned before being released and the distance (in minutes) from each of the 25 prisons in the system to their next of kin. 

Each of the prisons has on average has 230 of the prisoners in the dataset. The largest is the Chester prison, with 503 prisoners. The smallest is Greene with 101. On average, prisoners would move 139 minutes closer to home if they moved to their nearest prison. 

We use the distance from next of kin to the prison as an instrument, because previous research has established that when prisoners are closer to home they are more likely to be visited \citet{Jackson1997}. This approach assumes that distance to home has no other effect on recidivism not accounted for by observed covariates -- notably the effects of a particular prison's characteristics (prison fixed effects). Both of these are strong assumptions that ought to be examined. That being said, they are also both assumptions that would be made in a classic IV setting.

The intervention is then to shift prisoners both nearer and farther from home and estimate the impacts on recidivism among those for whom this shift changes their chance of being visited. This is a more plausible intervention than moving all prisoners to within a certain distance from home. Note, for example, that some prisoners' next of kin are a certain minimum distance from all Pennsylvania prisons. This means that any hypothesized intervention which moves them closer than that minimum distance is impossible. We measure distance in estimated travel time by car. The shift estimator therefore alleviates some of the issues with the positivity assumption, which requires that all subjects have positive probability of the theorized instrument value. The positivity assumption is not required by parametric models, and is strong. 

In our simulations, we used an unbounded $Z$. In this case, that no longer holds. No one can be a negative distance from prison, for example. We can easily incorporate this into our analysis, setting values for $z_{max}, z_{min}$ such that $Z \in [z_{min},z_{max}]$. In this dataset, we know the maximum and minimum distances prisoners can be from prison, and set $z_{max}, z_{min}$ accordingly. 

\begin{figure}[h!] 
	  \begin{center}
	    \includegraphics[width=.75\textwidth]{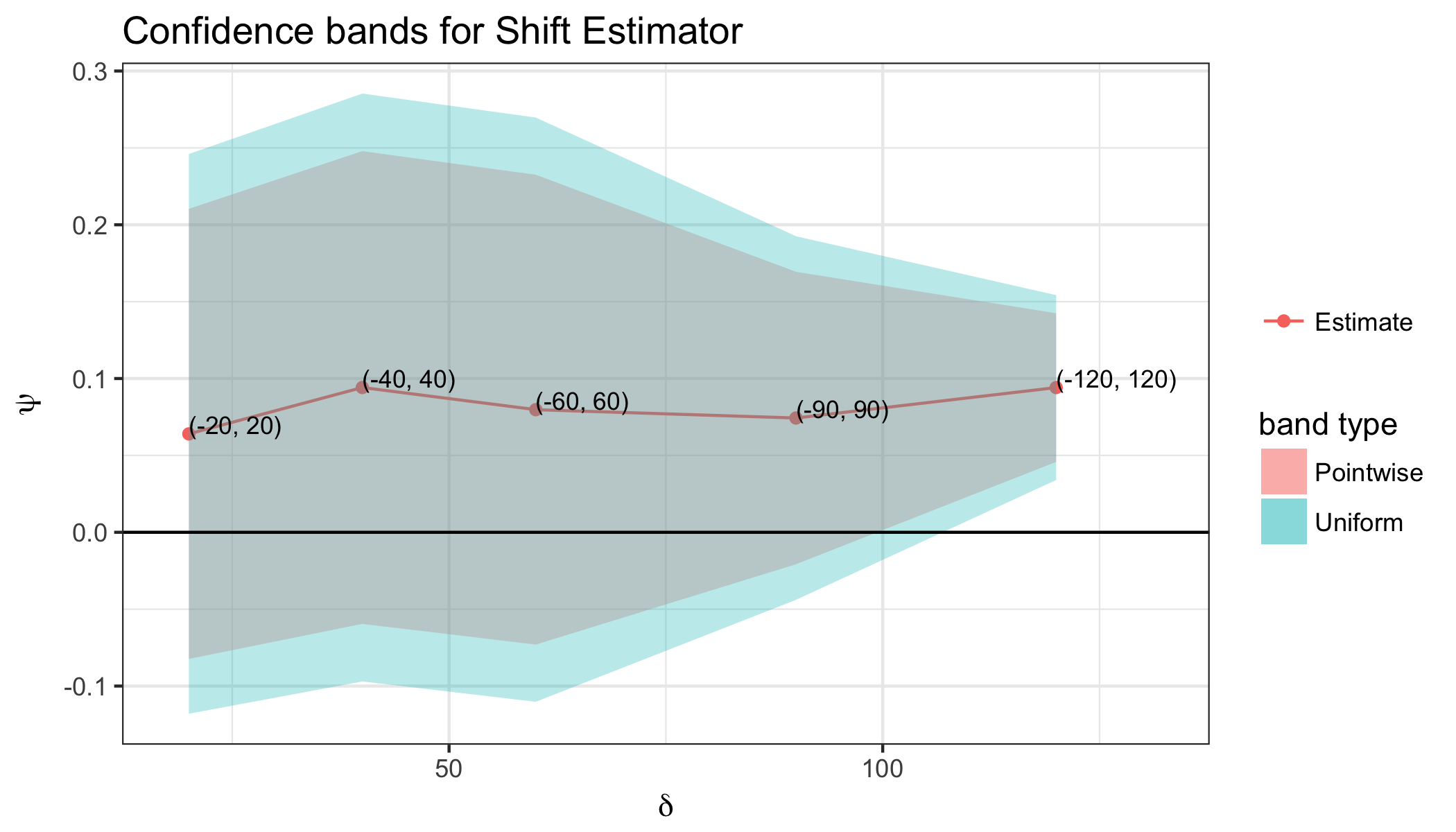}
	  \caption{Estimated Effects of Visitation Induced by Shifting Driving Distance on Recidivism}
	  \label{fig:DubShift}
	  \end{center}
\end{figure}

The results of shifts between 20 and 120 minutes nearer or further from home are illustrated in Figure \ref{fig:DubShift}. We run over 5 folds. In order to estimate the nuisance parameters, we use SuperLearner for the outcome regressions and ranger combined with a kernel for the conditional densities. See Section \ref{sec:Estimation} for details.

The estimated effect is an increase between 6.4\% to 9.4\%  in the chance of recidivating when visitation is lost. In this case, the complier population at some $\delta$ are those who are currently being visited, but would lose visitation if moved $\delta$ minutes farther away. It also includes those who are not being visited but would be if moved $\delta$ minutes closer. At a 120 minute shift levels, the 95\% pointwise and uniform confidence bands include no longer include zero, implying this is a statistically significant effect. The confidence bands narrow as the shifts get larger, likely due to the increased complier population. 

Another useful feature of the multiplier bootstrap is that we can test a hypothesis of effect homogeneity. To do so, we check whether we can draw a horizontal line across this figure which is fully contained within the confidence bounds. Here, we can do so at 0.25, for example, meaning we cannot reject the hypothesis of effect homogeneity.

The complier population ranges from about 6\% of prisoners at a 20 minute increase, to 25\% of prisoners at a 120 minute increase. 

\begin{figure}[h!] 
	  \begin{center}
	    \includegraphics[width=.75\textwidth]{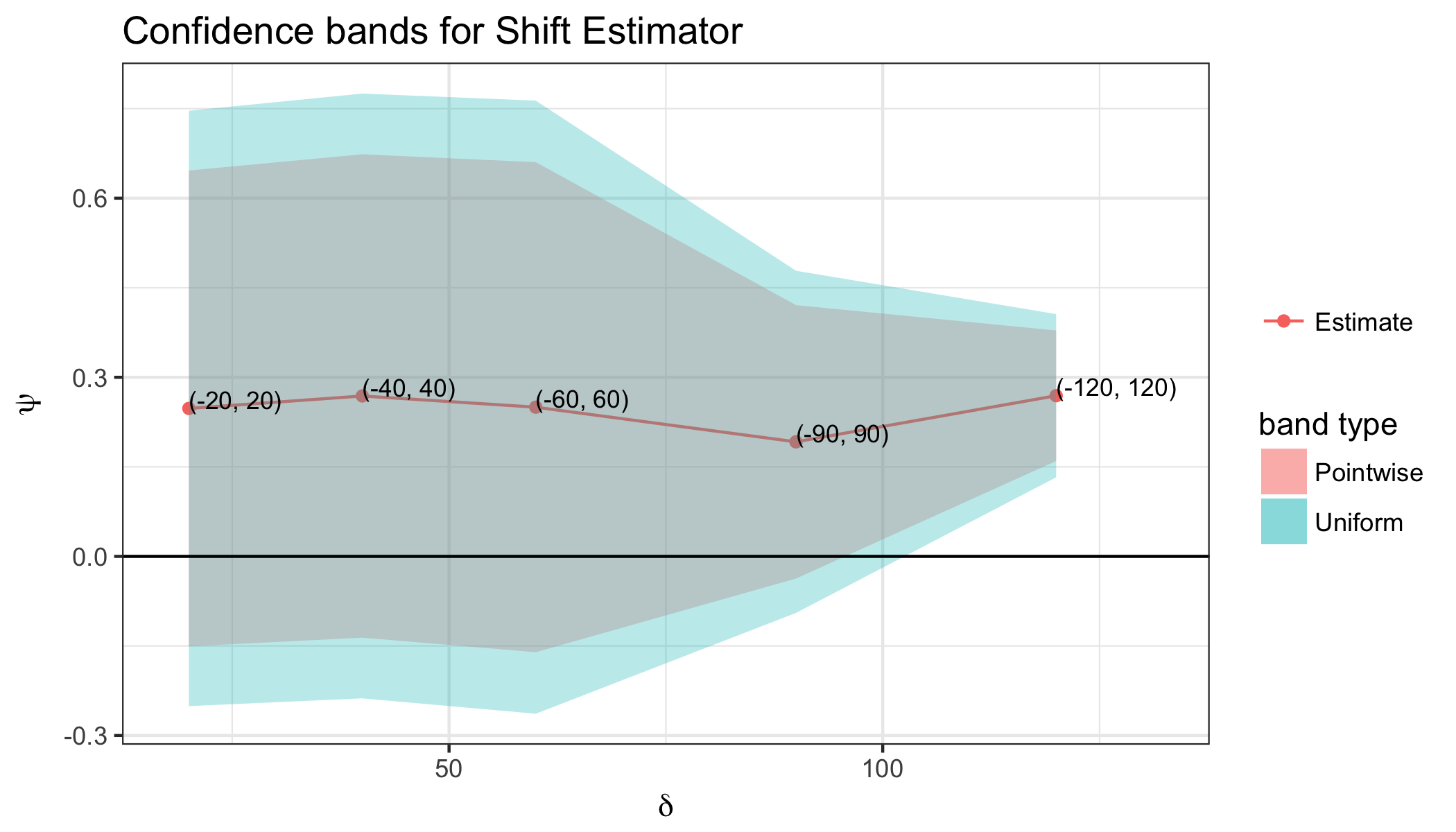}
	  \caption{Estimated Effects Child Visitation Induced by of a Positive Shift in Driving Distance on Recidivism}
	  \label{fig:DubShiftChild}
	  \end{center}
\end{figure}

These are the results for any visitation at all at the last location. We can similarly examine the effects of visitation among specific types of visitors. For example, we can specifically look at the effects of being visited by one's children. Note that although we do not restrict explicitly to inmates with children, the complier population should exclude anyone without children, since we necessarily cannot induce or end child visitation among the childless. The results of this analysis are given in Figure \ref{fig:DubShiftChild}.

These effects, again, are not significant at the 95\% level until a 120 minute shift. The point estimates are much higher than for visitation by anyone; they rise almost to 30\%. The complier population, on the other hand, is understandably much smaller. At the smallest shift level it is 2\% of the total population and rises to just 10\%. We again cannot reject homogeneity, since for example a 25\% effect is within the confidence bands at all $\delta$ levels. 

These results are summarized below:

\begin{table}[ht]
\caption{Regression Results}
\label{table:Regressions}
\centering
\begin{tabular}{lrrrrr}
$\delta$ value & 20 & 40 & 60 & 90 & 120 \\ 
\hline
\hline
All visits & 0.064 & 0.094 & 0.08 & 0.074 & 0.094 \\ 
Uniform CI & (-0.118,0.246) & (-0.097,0.285) & (-0.11,0.27) & (-0.044,0.193) & (0.034,0.154) \\ 
Compliers & 0.057 & 0.105 & 0.143 & 0.21 & 0.247 \\ 
\hline
Child visits & 0.248 & 0.269 & 0.25 & 0.192 & 0.269 \\ 
Uniform CI & (-0.251,0.746) & (-0.238,0.775) & (-0.264,0.763) & (-0.095,0.478) & (0.132,0.406) \\ 
Compliers & 0.023 & 0.045 & 0.06 & 0.087 & 0.096 \\ 
\hline\\
\end{tabular}
\end{table}

\section{Discussion and Conclusion}\label{sec:Discussion}
Our analyses provide evidence that increased visitation may reduce recidivism. The effects are potentially quite large, and given how difficult it is to reduce recidivism, should be encouraging to policy makers who have undertaken steps to keep prisoners near their families. Although the hypothesized intervention in this case was moving prisoners nearer to or farther from home, the results suggest that any intervention which increases visitation is likely to reduce recidivism.

Preliminary results also suggest there may be ways to target those inmates who are particularly sensitive to visitation. Especially in the analyses of child visitation, we saw that there may be effect heterogeneity. We hypothesize that those most affected are those with the strongest ties, but more analysis is needed to test that hypothesis.

There are some caveats to keep in mind. First, in our analysis, recidivism is measured as being rearrested for a new crime within three years. This is not a perfect measure of the act of recidivating. We seek to measure the effect of visitation on the overall probability of ever committing a new crime. Our measure, however, clearly overlooks anyone who is rearrested after three years. It also overlooks anyone who has committed a new crime, but has not been caught. Similarly, anyone who is falsely accused and convicted will be considered to have recidivated. Finally, those with a parole violation are not considered to have recidivated, but are also not free to recidivate after their rearrest (they are excluded from the analysis).
	
Second, we use distance as an instrument. This distance figure was reported by the prisoners upon entry into the prison system, and may not be entirely accurate. Worse, it may be biased in a way that is associated with recidivism. For example, if prisoners who are likely to commit more crimes tend to record a ``home'' location far from their prison. This would be an ignorability restriction, where there is some unobserved variable which links the distance metric and the probability of recidivating.

Third, it may be that prisoners are moved between prisons in a way that is associated with their chances of committing a new crime. If prisoners are moved closer to home because of good behavior, those who are less likely to commit new crimes are also likely to be closer to home. This is another violation of the ignorability assumption. 

There may additionally be a causal path between distance and recidivism that does not pass through the treatment or visitation. Perhaps prisoners feel more comfortable when closer to home and suffer less stress. Conversely, it may be that  their criminal networks are more easily maintained close to home, making them more likely to recidivate. These are both examples of exclusion restriction violations.

Depending on the way we hypothesize a change in distance, there may also be positivity violations. These occur when a prisoner cannot receive the instrument value we assign them. Monotonicity, instrumentation, and consistency are probably unlikely to be violated in this example.

The need for robust causal inference tools for observational data is pressing. In this paper, we have presented one such tool for new estimands based on dynamic interventions on instrumental variables. These estimators allow researchers to investigate interventions that consider the impact of each individual having their observed instrument value perturbed, rather than setting the instrument to a specific level, which can require strong and impractical assumptions. Our tools also allow researchers to make use of nonparametric machine learning tools, while still allowing fast parametric-rate inference.

\section*{acknowledgements}
The authors would like to thank the Pennsylvania Department of Corrections for providing the data. 


\bibliography{refs2}

\begin{thebibliography}{39}
\expandafter\ifx\csname natexlab\endcsname\relax\def\natexlab#1{#1}\fi
\expandafter\ifx\csname url\endcsname\relax
  \def\url#1{\texttt{#1}}\fi
\expandafter\ifx\csname urlprefix\endcsname\relax\def\urlprefix{URL: }\fi

\bibitem[{Acemoglu(2002)}]{Acemoglu2002}
Acemoglu, D. (2002) {Technical Change , and the Labor Inequality , Market}.
\newblock \textit{Journal of Economic Literature}, \textbf{40}, 7--72.

\bibitem[{Angrist et~al.(1996)Angrist, Imbens and Rubin}]{Angrist1996}
Angrist, J.~D., Imbens, G.~W. and Rubin, D.~B. (1996) {Identification of Causal
  Effects Using Instrumental Variables}.
\newblock \textit{Journal of the American Statistical Assocation}, \textbf{91},
  444 -- 472.

\bibitem[{Bales and Mears(2008)}]{Bales2008}
Bales, W.~D. and Mears, D.~P. (2008) {Inmate social ties and the transition to
  society: Does visitation reduce recidivism?}
\newblock \textit{Journal of Research in Crime and Delinquency}, \textbf{45},
  287 -- 321.

\bibitem[{Bersani and Doherty(2018)}]{Bersani2018}
Bersani, B.~E. and Doherty, E.~E. (2018) {Desistance from Offending in the
  Twenty-First Century}.
\newblock \textit{Annual Review of Criminology}, \textbf{1}, 311--334.
\newblock
  \urlprefix\url{https://doi.org/10.1146/annurev-criminol-032317-092112.%0Ahttp://www.annualreviews.org/doi/pdf/10.1146/annurev-criminol-032317-092112}.

\bibitem[{Bickel et~al.(1975)Bickel, Hammel and O'Connell}]{Bickel1975}
Bickel, P.~J., Hammel, E.~A. and O'Connell, J.~W. (1975) {Sex bias in graduate
  admissions: Data from Berkeley}.
\newblock \textit{Science}, \textbf{187}, 398--404.

\bibitem[{Borgman(1985)}]{Borgman1985}
Borgman, R. (1985) {The influence of family visiting upon boys' behavior in a
  juvenile correctional institution.}
\newblock \textit{Child welfare}, \textbf{54}, 629--638.

\bibitem[{Cain et~al.(2010)Cain, Robins, Lanoy, Logan, Costagliola and
  Hern{\'{a}}n}]{Cain2010}
Cain, L.~E., Robins, J.~M., Lanoy, E., Logan, R., Costagliola, D. and
  Hern{\'{a}}n, M.~A. (2010) {When to Start Treatment? A Systematic Approach to
  the Comparison of Dynamic Regimes Using Observational Data}.
\newblock \textit{The International Journal of Biostatistics}, \textbf{6}.
\newblock
  \urlprefix\url{https://www.degruyter.com/view/j/ijb.2010.6.2/ijb.2010.6.2.1212/ijb.2010.6.2.1212.xml}.

\bibitem[{Casey-Acevedo et~al.(2004)Casey-Acevedo, Bakken and
  Karle}]{Casey-Acevedo2004}
Casey-Acevedo, K., Bakken, T. and Karle, A. (2004) {Children visiting mothers
  in prison: The effects on mothers' behaviour and disciplinary adjustment}.
\newblock \textit{Australian and New Zealand Journal of Criminology},
  \textbf{37}, 418--430.

\bibitem[{Cochran(2014)}]{Cochran2014}
Cochran, J.~C. (2014) {Breaches in the Wall: Imprisonment, Social Support, and
  Recidivism}.
\newblock \textit{Journal of Research in Crime and Delinquency}, \textbf{51},
  200--229.

\bibitem[{Derkzen et~al.(2009)Derkzen, Gobeil and Gileno}]{Derkzen2009}
Derkzen, D., Gobeil, R. and Gileno, J. (2009) {Visitation and Post-Release
  Outcome Among Federally-Sentenced Offenders}.
\newblock \textit{Tech. rep.}, Correctional Service of Canada, Ottowa, Ontario.

\bibitem[{D{\'{i}}az~Mu{\~{n}}oz and van~der Laan(2011)}]{Diaz2011}
D{\'{i}}az~Mu{\~{n}}oz, I. and van~der Laan, M.~J. (2011) {Super Learner Based
  Conditional Density Estimation with Application to Marginal Structural
  Models}.
\newblock \textit{The International Journal of Biostatistics}, \textbf{7},
  1--20.
\newblock
  \urlprefix\url{https://www.degruyter.com/view/j/ijb.2011.7.issue-1/1557-4679.1356/1557-4679.1356.xml}.

\bibitem[{Dudik et~al.(2014)Dudik, Langford and Li}]{Dudik2014}
Dudik, M., Langford, J. and Li, L. (2014) {Doubly Robust Policy Evaluation and
  Learning}.
\newblock \textit{Statistical Science}, \textbf{29}, 485--511.
\newblock \urlprefix\url{http://arxiv.org/abs/1103.4601}.

\bibitem[{Durose et~al.(2014)Durose, Cooper and Snyder}]{Durose2014}
Durose, M.~R., Cooper, A.~D. and Snyder, H.~N. (2014) {Recidivism of Prisoners
  Released in 30 States in 2005: Patterns from 2005 to 2010}.
\newblock \textit{Tech. rep.}, US Department of Justice, Office of Justice
  Programs, Bureau of Justice Statistics.

\bibitem[{Duwe and Clark(2013)}]{Duwe2013}
Duwe, G. and Clark, V. (2013) {Blessed Be the Social Tie That Binds: The
  Effects of Prison Visitation on Offender Recidivism}.
\newblock \textit{Criminal Justice Policy Review}, \textbf{24}, 271--296.

\bibitem[{Etil{\'{e}} and Sharma(2015)}]{Etile2015}
Etil{\'{e}}, F. and Sharma, A. (2015) {Do High Consumers of Sugar-Sweetened
  Beverages Respond Differently to Price Changes? A Finite Mixture IV-Tobit
  Approach}.
\newblock \textit{Health Economics (United Kingdom)}, \textbf{24}, 1147--1163.

\bibitem[{Haneuse and Rotnitzky(2013)}]{Haneuse2013}
Haneuse, S. and Rotnitzky, A. (2013) {Estimation of the effect of interventions
  that modify the received treatment}.
\newblock \textit{Statistics in Medicine}, \textbf{32}, 5260--5277.

\bibitem[{Heckman(1990)}]{Heckman1990}
Heckman, J. (1990) {Varieties of Selection Bias}.
\newblock \textit{The American Economic Review}, \textbf{80}, 313--318.
\newblock \urlprefix\url{http://www.jstor.org/stable/2006591}.

\bibitem[{Holt and Miller(1972)}]{Holt1972}
Holt, N. and Miller, D. (1972) {Explorations in Inmate-Family Relationships}.
\newblock \textit{Tech. rep.}, Research Division, Californa Department of
  Corrections.

\bibitem[{Izbicki and Lee(2017)}]{Izbicki2016}
Izbicki, R. and Lee, A.~B. (2017) {Converting High-Dimensional Regression to
  High-Dimensional Conditional Density Estimation}.
\newblock \textit{arXiv preprint}, 1--43.

\bibitem[{Izbicki et~al.(2014)Izbicki, Lee and Schafer}]{Izbicki2014}
Izbicki, R., Lee, A.~B. and Schafer, C.~M. (2014) {High-Dimensional Density
  Ratio Estimation with Extensions to Approximate Likelihood Computation}.
\newblock In \textit{17th International Conferenc on Artificial Intelligence
  and Statistics (AISTATS)}, vol.~33. Reykjavik, Iceland: JMLR: W {\&} CP.

\bibitem[{Jackson et~al.(1997)Jackson, Templer, Reimer and
  LeBaron}]{Jackson1997}
Jackson, P., Templer, D.~I., Reimer, W. and LeBaron, D. (1997) {Correlates of
  Visitation in a Men's Prison}.
\newblock \textit{International Journal of Offender Therapy and Comparative
  Criminology}, \textbf{41}, 79--85.
\newblock \urlprefix\url{https://doi.org/10.1177/0306624X9704100108}.

\bibitem[{Kennedy(2017)}]{kennedy2017semiparametric}
Kennedy, E.~H. (2017) Semiparametric theory.
\newblock \textit{Wiley StatsRef: Statistics Reference Online}, 1--7.

\bibitem[{Kennedy(2019)}]{kennedy2019nonparametric}
--- (2019) Nonparametric causal effects based on incremental propensity score
  interventions.
\newblock \textit{Journal of the American Statistical Association},
  \textbf{114}, 645--656.

\bibitem[{van~der Laan and Petersen(2007)}]{VanDerLaan2007}
van~der Laan, M.~J. and Petersen, M.~L. (2007) {Causal Effect Models for
  Realistic Individualized Treatment and Intention to Treat Rules}.
\newblock \textit{International Journal of Biostatistics}, \textbf{3}, 1--51.

\bibitem[{van~der Laan and Robins(2003)}]{VanderLaan2003}
van~der Laan, M.~J. and Robins, J.~M. (2003) \textit{{Unified Methods for
  Censored Longitudinal Data and Causality}}.
\newblock \urlprefix\url{http://link.springer.com/10.1007/978-0-387-21700-0}.

\bibitem[{Mears et~al.(2012)Mears, Cochran, Siennick and Bales}]{Mears2012}
Mears, D.~P., Cochran, J.~C., Siennick, S.~E. and Bales, W.~D. (2012) {Prison
  Visitation and Recidivism}.
\newblock \textit{Justice Quarterly}, \textbf{29}, 889 -- 918.

\bibitem[{Moore et~al.(2012)Moore, Neugebauer, Van~der Laan and
  Tager}]{Moore2012}
Moore, K.~L., Neugebauer, R., Van~der Laan, M.~J. and Tager, I.~B. (2012)
  {Causal inference in epidemiological studies with strong confounding}.
\newblock \textit{Statistics in Medicine}, \textbf{31}, 1380--1404.

\bibitem[{Munoz and Van Der~Laan(2012)}]{Munoz2012}
Munoz, I.~D. and Van Der~Laan, M.~J. (2012) {Population Intervention Causal
  Effects Based on Stochastic Interventions}.
\newblock \textit{Biometrics}, \textbf{68}, 541--549.

\bibitem[{Murphy(2003)}]{Murphy2003}
Murphy, S. (2003) {Optimal Dynamic Treatment Regimes}.
\newblock \textit{Journal of the Royal Statistical Society. Series B
  (Statistical Methodology)}, \textbf{65}, 331--366.
\newblock
  \urlprefix\url{http://www.jstor.org/stable/3647509%5Cnpapers2://publication/uuid/F3ADCD88-418C-4685-8BEA-D5B12B72C65C}.

\bibitem[{Pearl(2009)}]{Pearl2009}
Pearl, J. (2009) \textit{{Causality: Models, Reasoning, {\&} Inference}}.
\newblock Cambridge, MA: Cambridge University Press.

\bibitem[{Robins(2004)}]{Robins2004}
Robins, J.~M. (2004) {Optimal Structural Nested Models for Optimal Sequential
  Decisions}.
\newblock In \textit{Proceedings of the Second Seattle Symposium in
  Biostatistics. Lecture Notes in Statistics.}, vol. 179, 189--326. New York,
  NY: Springer.
\newblock
  \urlprefix\url{http://link.springer.com/10.1007/978-1-4419-9076-1_11}.

\bibitem[{Robins et~al.(2008)Robins, Orellana and Rotnitzky}]{Robins2008}
Robins, J.~M., Orellana, L. and Rotnitzky, A. (2008) {Growth rates in epidemic
  models: application to a model for HIV/AIDS progression}.
\newblock \textit{Statistics in medicine}, \textbf{27}, 4678--4721.

\bibitem[{Rubin(1974)}]{rubin1974}
Rubin, D.~B. (1974) {Estimating causal effects of treatments in randomized and
  nonrandomized studies}.
\newblock \textit{Journal of educational Psychology}, \textbf{66}, 688--701.

\bibitem[{Taubman et~al.(2009)Taubman, Robins, Mittleman and
  Hern{\'{a}}n}]{Taubman2009}
Taubman, S.~L., Robins, J.~M., Mittleman, M.~A. and Hern{\'{a}}n, M.~A. (2009)
  {Intervening on risk factors for coronary heart disease: An application of
  the parametric g-formula}.
\newblock \textit{International Journal of Epidemiology}, \textbf{38},
  1599--1611.

\bibitem[{Tian(2008)}]{Tian2008}
Tian, J. (2008) {Identifying Dynamic Sequential Plans}.
\newblock \textit{Uncertainty in Artificial Intelligence}, 554--561.
\newblock
  \urlprefix\url{http://uai.sis.pitt.edu/papers/08/p554-tian.pdf%5Cnhttp://uai2008.cs.helsinki.fi/UAI_camera_ready/tian.pdf}.

\bibitem[{Tsiatis(2006)}]{Tsiatis2006}
Tsiatis, A.~A. (2006) \textit{{Semiparametric Theory and Missing Data}}.
\newblock Raleigh, NC: Springer Series in Statistics.

\bibitem[{van~der Vaart(1998)}]{vanderVaart1998AsymptoticStatistics}
van~der Vaart, A. (1998) \textit{{Asymptotic Statistics}}.
\newblock Cambridge, MA: Cambridge University Press.

\bibitem[{Wooldridge(1999)}]{Wooldridge1999}
Wooldridge, J.~M. (1999) {Distribution-free estimation of some nonlinear panel
  data models}.
\newblock \textit{Journal of Econometrics}, \textbf{90}, 77--97.

\bibitem[{Young et~al.(2014)Young, Hern{\'{a}}n and Robins}]{Young2014}
Young, J.~G., Hern{\'{a}}n, M.~A. and Robins, J.~M. (2014) {Identification,
  Estimation and Approximation of Risk under Interventions that Depend on the
  Natural Value of Treatment Using Observational Data}.
\newblock \textit{Epidemiol. Methods}, \textbf{3}, 1--19.

\end{thebibliography}

\null\newpage

\section{Appendix}\label{sec:Appendix}

\subsection{Comparison to Two-Stage Least Squares}
Two stage least squares assumes a parametric model and constant effects, among other things. Thus, by assumption, the estimate of the causal effect of visitation on recidivism is the Average Treatment Effect. Here, we estimate that losing visitation causes the rate of recidivism to rise by about 10\% for everyone in our population. This is not significantly different from 0 according to this model. The 10\% figure is not directly comparable to the estimates we have found, since our estimates are local effects, and this is a global effect. That being said, at our largest complier population, we find about a 11\% effect of losing visitation, so the two are not wildly different. 

\begin{table}[ht!]
\centering
\begin{tabular}{rrrrr}
  \hline
 & Estimate & Std. Error & t value & Pr(>|t|) \\ 
  \hline
(Intercept) & 0.32 & 0.08 & 4.22 & 0.00 \\ 
  No Visits at Last Location & 0.10 & 0.06 & 1.53 & 0.13 \\ 
  White & -0.01 & 0.02 & -0.67 & 0.50 \\ 
  Length of Stay Last Location & 0.00 & 0.00 & 1.18 & 0.24 \\ 
  County 3& -0.05 & 0.02 & -3.39 & 0.00 \\ 
  County 4& 0.02 & 0.05 & 0.39 & 0.70 \\ 
  County 5& -0.00 & 0.05 & -0.10 & 0.92 \\ 
  County 6& -0.01 & 0.05 & -0.23 & 0.82 \\ 
  County 7& 0.01 & 0.08 & 0.08 & 0.94 \\ 
  County 8 & -0.08 & 0.10 & -0.88 & 0.38 \\ 
  Age & -0.01 & 0.00 & -10.66 & 0.00 \\ 
  Urban & 0.05 & 0.05 & 1.09 & 0.28 \\ 
  Prior Arrests & 0.01 & 0.00 & 10.91 & 0.00 \\ 
  Married & 0.01 & 0.02 & 0.31 & 0.75 \\ 
  Violent & 0.01 & 0.02 & 0.92 & 0.36 \\ 
  LSIR Score & 0.00 & 0.00 & 4.18 & 0.00 \\ 
  HS Grad & 0.01 & 0.01 & 1.08 & 0.28 \\ 
  Custody Level & 0.06 & 0.01 & 7.87 & 0.00 \\ 
  Num of Prior Inc & 0.04 & 0.01 & 5.63 & 0.00 \\ 
  Num of total misconducts & 0.00 & 0.00 & 1.46 & 0.15 \\ 
   \hline
\end{tabular}
\end{table}

\null\newpage

\subsection{Additional Simulations}

We can see how the convergence behaves as we change the size of simulated dataset. At 100, there is a great deal of noise and the IF-based estimator is unstable. By 1,000, however, it seems to be stabilized. At 10,000 the estimates are very exact.
\begin{figure}[ht!]
    \centering
    \includegraphics[width = \textwidth]{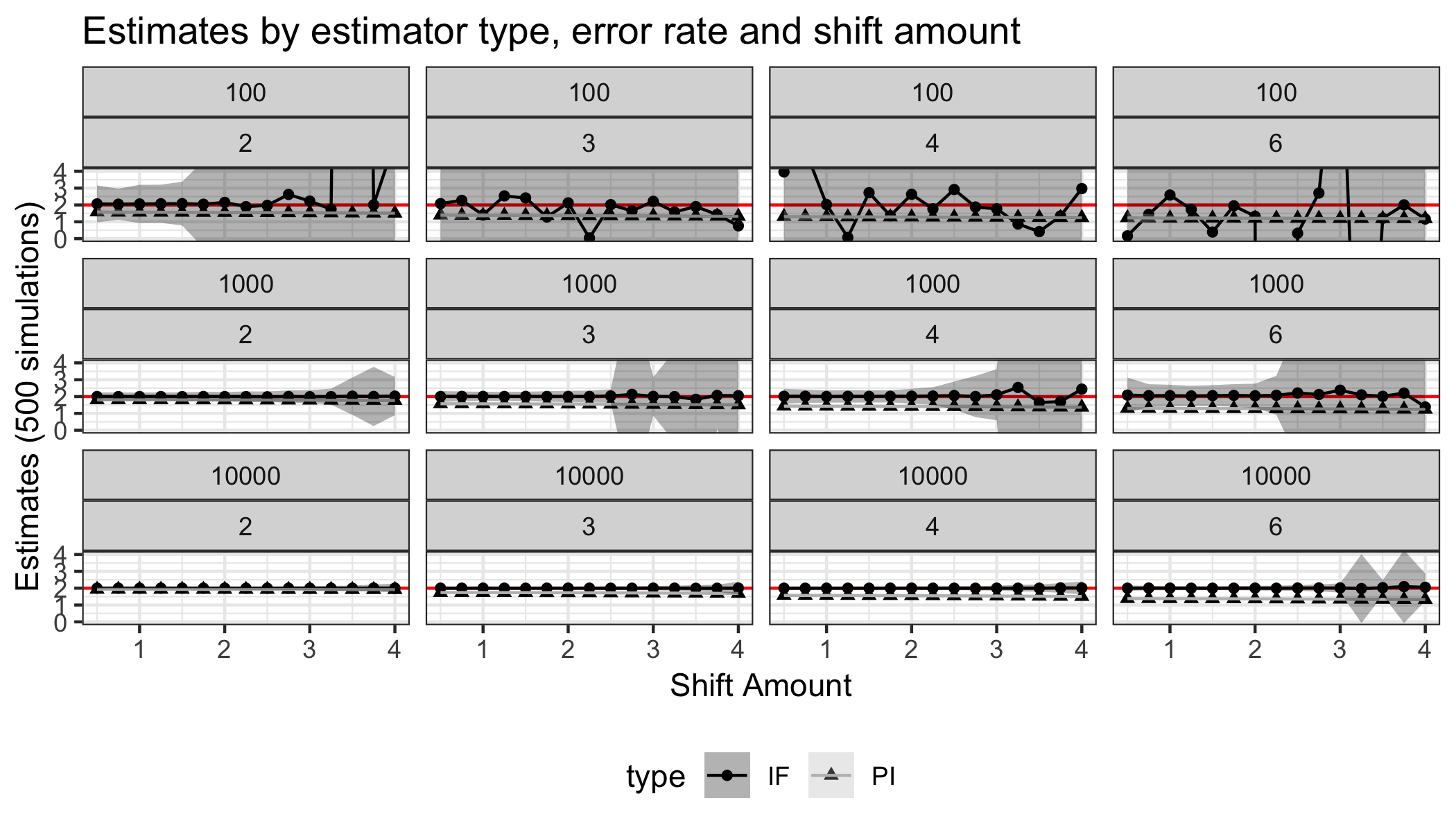}
    \caption{Simulations at 100, 1,000 and 10,000 sample sizes}
    \label{fig:sim100}
\end{figure}

We can also look at these figures taking the standard deviation we would use in practice, derived from the influence function. These are generally though not necessarily, slightly narrower.
\begin{figure}[ht!]
    \centering
    \includegraphics[width = .5\textwidth]{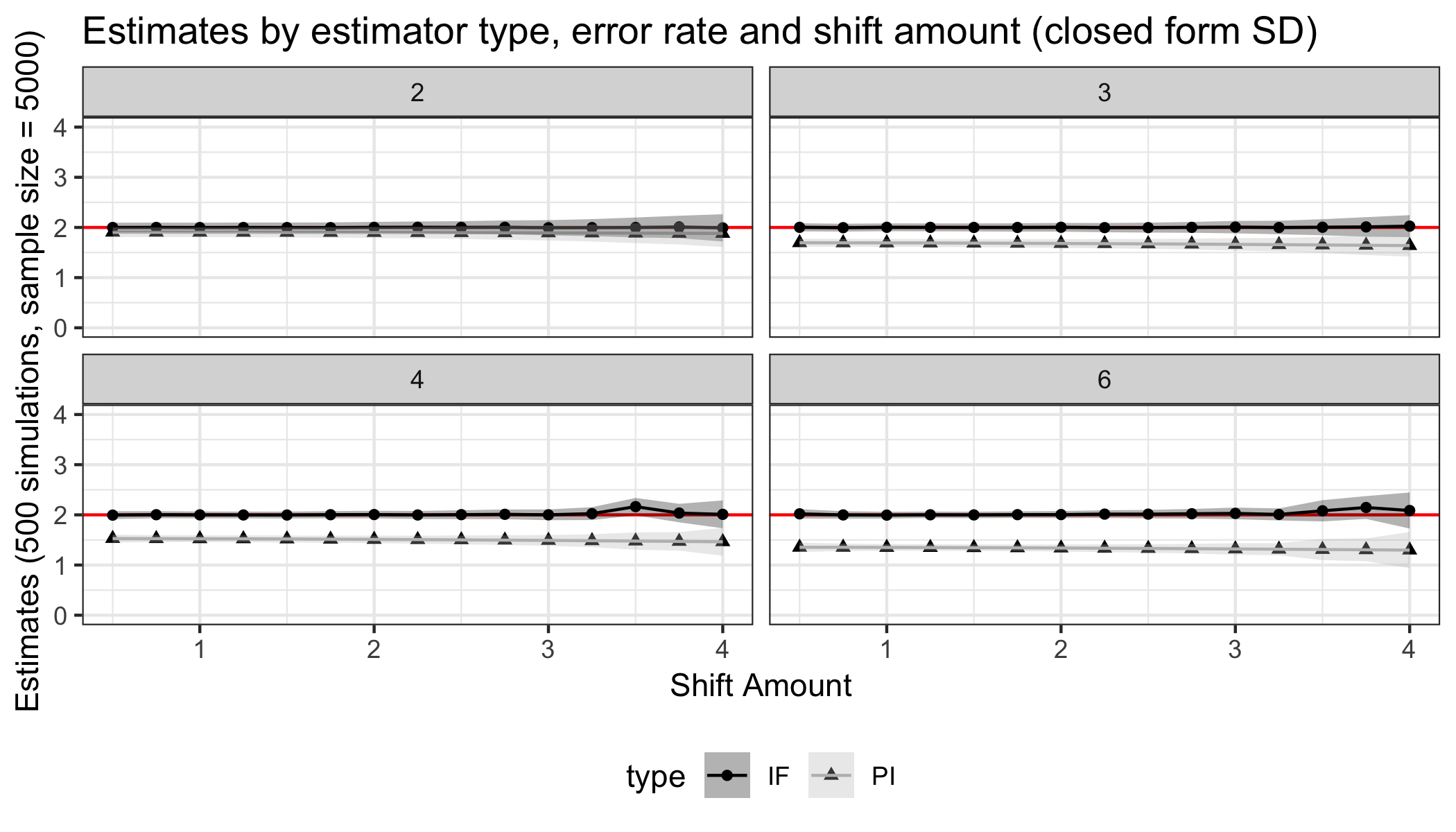}
    \caption{Simulations showing the confidence intervals derived from closed-form SD}
    \label{fig:sim100cf}
\end{figure}

\subsection{Estimation}\label{sec:Estimation}
Our influence-function based estimator leaves the user a choice of how to estimate the nuisance parameters. The trade off is generally between flexibility and precision. If the user is willing to assume the nuisance functions take parametric forms, estimates may appear more precise (have narrower confidence bands). However, if the user has misspecified the parametric form, these confidence bands may be misleading. 

In our results, we use SuperLearner, an ensemble learner, to estimate the regression functions. Our library includes the following learners: \texttt{"SL.gam", "SL.glm", "SL.glm.interaction", "SL.glm", "SL.ranger", and "SL.mean"}. 

To estimate the conditional densities, we adapt a kernel approach. First, we model the mean $\mu_z = \expt{Z \mid \bX}$ using ranger (a version of random forests). We find the variance as 

\begin{align}
	\sigma_z^2 = \expt{ \left[Z - \mathbb{E}(Z \mid \bX) \right]^2  }
\end{align}

We then estimate the conditional density as

\begin{align}
	f(z\mid \bx) = K\left( \frac{z - \hat{\mu}_z}{\hat{\sigma}_z^2} \right)
\end{align}

Where $K$ is a Gaussian kernel. 

\begin{figure}[ht!]
	\centering
	\includegraphics[width = .5\textwidth]{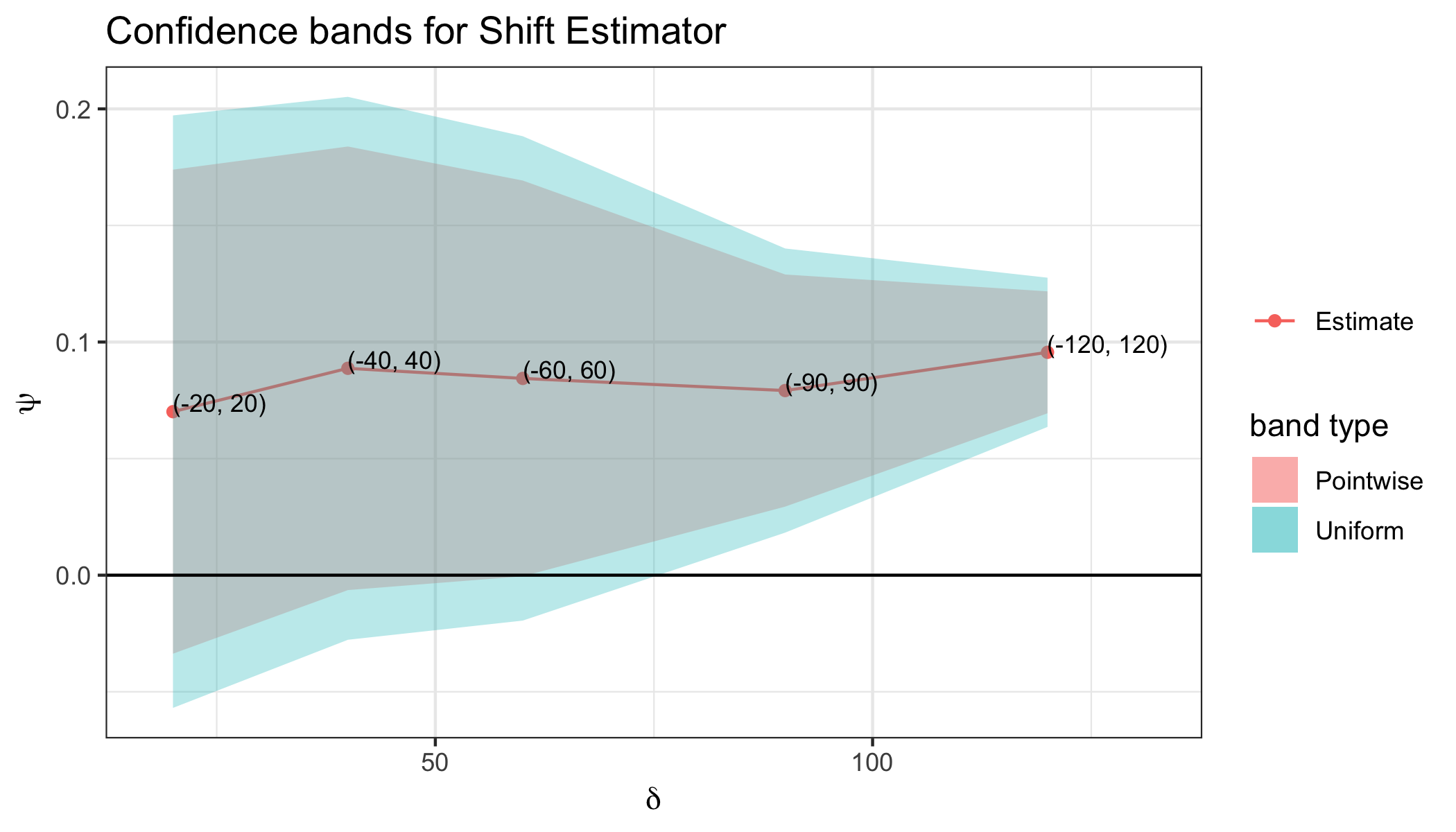}
	\caption{Main analysis repeated with GLM as the estimator of the propensity score mean function}
	\label{fig:doubleGLM}
\end{figure}

This gives flexible but relatively imprecise estimates of the conditional density. If we replace ranger as the estimator of the mean function with glm, our final estimates of the confidence intervals narrow considerably. We repeat the analysis using a glm estimate of the mean of the propensity score function, and see in Figure \ref{fig:doubleGLM} that the point estimates are largely unchanged, but the confidence bands narrow.

\begin{figure}[ht!]
	\centering
	\includegraphics[width = .5\textwidth]{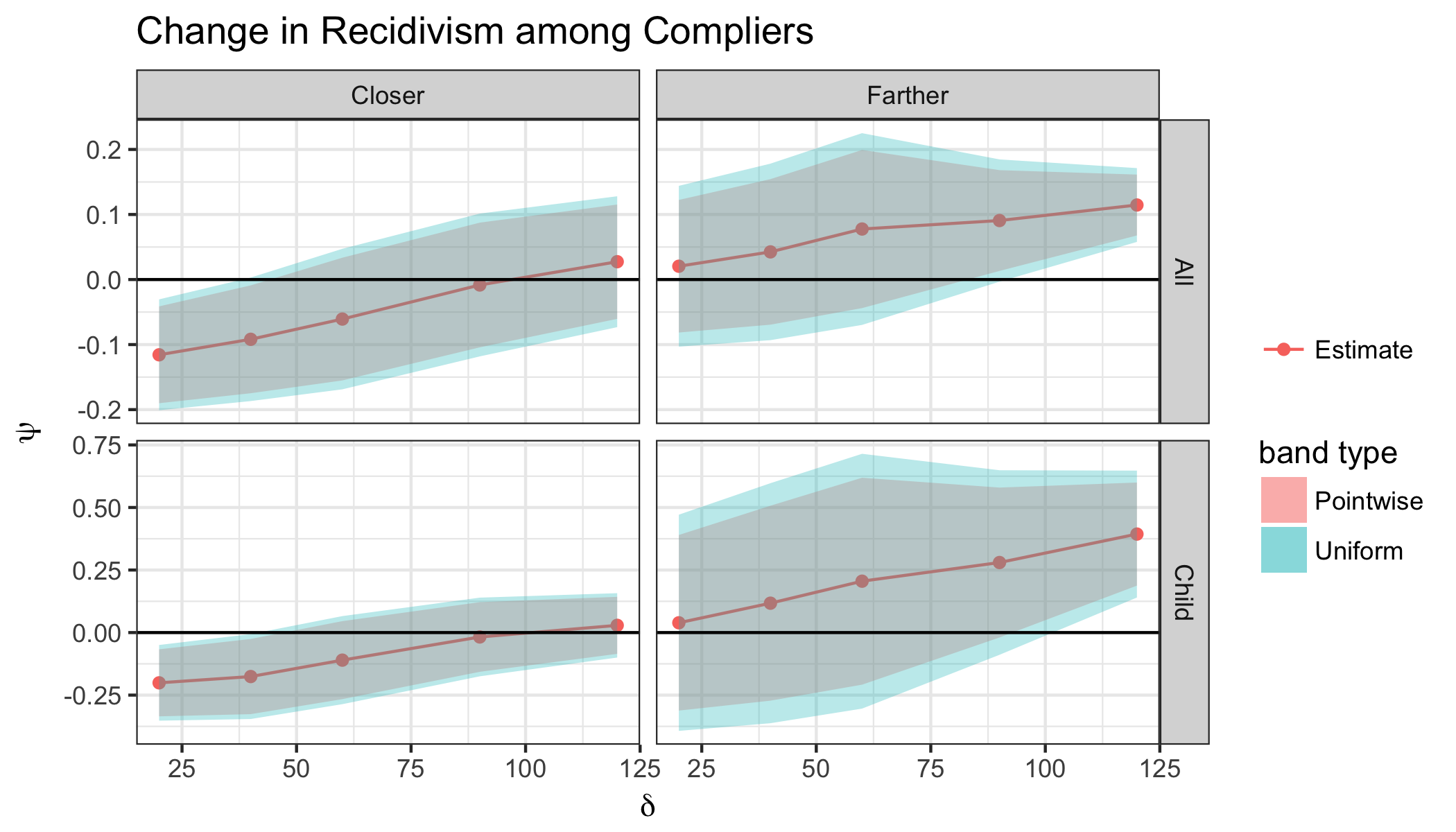}
	\caption{Results for a single shift intervention when conditional densities are estimated using glm and a kernel rather than in a fully nonparametric fashion}
	\label{fig:recidAllglm}
\end{figure}

The results of the analysis of a single shift estimate using GLM are given in Figure \ref{fig:recidAllglm}. In this case, the uniform confidence bands do not always include 0, so the choice of nuisance parameter estimator can make a considerable difference to the conclusions we draw. 

Future work on these continuous IV estimators should incorporate refinement of the conditional density estimation. 

In fact, work in the area of nonparametric conditional density estimation is advancing rapidly. Izbicki, Lee and colleagues (\citep{Izbicki2014, Izbicki2016}, eg) proposes converting conditional density estimation into a problem of many nonparametric regression functions. Their package, \texttt{FlexCode}, is an implementation of this work. Likewise, Ivan Diaz and Mark van der Laan have developed methods for nonparametric conditional density estimation \cite{Diaz2011}. This is implemented in their \texttt{condensier} package. In our case, these approaches resulted in unstable values for the conditional density estimates, but we believe they hold great promise. 

Another promising line of research is to target the ratio of density estimates directly. The lab of Masashi Sugiyama at the University of Tokyo has made considerable inroads in this area. Unfortunately, these are not implemented in R yet. Again, we believe this line of research holds enormous potential. 

\null\newpage

\subsection{Proofs}
\subsubsection{Proof of identification}
\begin{proof}
	We will show that 
	\begin{align}
		\psi_{shift} &= \mathbb{E}\{ Y^{A=1} - Y^{A=0} |  A^{Z+\delta}>A^{Z-\delta} \}\\
		&=\frac{\expt{\expt{Y|Z+\delta,X} - \expt{Y|Z-\delta,X} } }{\expt{\expt{A|Z+\delta,X} - \expt{A|Z-\delta,X} } }
	\end{align}
	
	This follows from (slightly modified forms of) the usual assumptions made in instrumental variables cases:
	\begin{enumerate}
		\item Consistency: $Z= z \Rightarrow A = A^z$ and $A = a \Rightarrow Y = Y^{a}$
		\item Ignorability: $Z \ci (Y^{z}, A^{z})|\bX$ 
		\item Exclusion Restriction: $Y^{za} = Y^a$
		\item Monotonicity: $\frac{d}{dz}A^z \geq 0$ a.e. or $\frac{d}{dz}A^z \leq 0$ a.e.
		\item Instrumentation: $\prob{A^{Z+\delta}>A^{Z}} >0$
		\item Positivity: $\prob{\epsilon < \prob{Z\pm \delta \mid \bX} < 1-\epsilon} = 1$
	\end{enumerate}
	Recall that we are using $Y^a$ to denote the potential outcome $Y^{a{Z^a}}$. The above assumptions allow us to make the following statements.
	
	\begin{align}
		\expt{Y^{1} - Y^{0} | A^{Z+\delta} > A^{Z-\delta}} &= \expt{Y^{1} - Y^{0} \mathbb{I}\{A^{Z+\delta} > A^{Z-\delta}\}}/ \prob{A^{Z+\delta} > A^{Z-\delta}} \n
		\prob{A^{Z+\delta} > A^{Z-\delta}} &= \prob{A^{Z+\delta}=1, A^{Z-\delta}=0}  \n
		&= \expt{\mathbb{I}\{A^{Z+\delta}=1, A^{Z-\delta}=0\}}  \n
		&= \expt{\expt{\mathbb{I}\{A^{Z+\delta}=1, A^{Z-\delta}=0\}|X}}  \n
		&= \expt{ \expt{A^{Z+\delta} - A^{Z-\delta}|X} } \n
		&= \expt{ \expt{A|Z+\delta,X} - \expt{A|Z-\delta,X} } \n
		\expt{(Y^{1} - Y^{0}) \mathbb{I}\{A^{Z+\delta}>A^{Z-\delta}\} } &= \expt{\expt{(Y^{1} - Y^{0}) \mathbb{I}\{A^{Z+\delta}>A^{Z-\delta}\}|X}} \n
		&= \expt{\expt{Y|Z+\delta,X} - \expt{Y|Z-\delta,X} } 
	\end{align}
	Where the final line follows because $Y^{Z=i} = Y^{Z=i,A^i} = Y^{A^i}$. This gives the desired result:
	
	\begin{align}
		\psi_{shift} &= \frac{\expt{\expt{Y|Z+\delta,X} - \expt{Y|Z-\delta,X} } }{\expt{\expt{A|Z+\delta,X} - \expt{A|Z-\delta,X} } }
	\end{align}
\end{proof}

\subsubsection{General derivation of the Influence Function takes the form given}\label{sec:EIF}
\begin{proof}
	Recall that our parameter of interest is:
	\begin{align}
		\label{parameter2}
		\psi_{shift} &= \frac{\expt{\expt{Y|Z+\delta,\bX} - \expt{Y|Z-\delta,\bX} } }{\expt{\expt{A|Z+\delta,\bX} - \expt{A|Z-\delta,\bX} } }\\
		&= \frac{\expt{\expt{Y|Z+\delta,\bX}} - \expt{\expt{Y|Z-\delta,\bX} } }{\expt{\expt{A|Z+\delta,\bX}} - \expt{\expt{A|Z-\delta,\bX} } }
	\end{align}
	
	To find the efficient influence function (IF) of this quantity, first we find the IF of each piece:
	\begin{enumerate}
		\item $\expt{Y|Z+\delta,\bX} = \mu(Z+\delta, \bX)$
		\item $\expt{Y|Z,\bX}= \mu(Z-\delta, \bX)$
		\item $\expt{A|Z+\delta,\bX}= \lambda(Z+\delta, \bX)$
		\item $\expt{A|Z,\bX}=\lambda(Z-\delta,\bX)$
	\end{enumerate}
	
	An influence function in the discrete case is a particular Gateaux derivative in the direction of a point mass. In the continuous case, it is an approximation of this derivative. Because of this, we can use the usual derivative rules when dealing with IF's, notably:
	\begin{align}\label{eq:chainrule}
		IF\left(\frac{f}{g}\right) &= \frac{IF(f)\cdot g - IF(g)\cdot f}{g^2}
	\end{align}
	
	Below we give the derivation of the IF for quantity (1). The notation $Z^{obs}$ denotes the observed value of $Z$, and likewise for $X$.
	\begin{align}
		IF[\mu(Z+\delta,  \bX)] &= IF \left[ \int \expt{Y \mid \bX = \bx, Z = z+\delta} d\prob{Z=z\mid \bX = \bx}d\prob{\bX = \bx} \right]\\
		&= \sum_x \sum_z \frac{\ic{\bX^{obs} = \bx}\ic{Z^{obs} = z+\delta}}{\prob{Z=z+\delta \mid \bX}\prob{\bX=\bx}} \left[Y - \expt{Y\mid \bX = \bx, Z=z+\delta}\right]\prob{Z=z\mid \bX}\prob{\bX = \bx}\\
		&\ \ + \sum_x \sum_z \expt{Y \mid \bX = \bx, Z = z+\delta} \frac{\ic{\bX^{obs}=\bx}}{\prob{\bX = \bx}}\left[\ic{Z^{obs} = z} - \prob{Z=z \mid \bX = \bx}\right] \prob{\bX = \bx}\\
		&\ \ + \sum_x \sum_z \expt{Y \mid \bX = \bx, Z = z+\delta} \prob{Z = z\mid \bX = \bx} \left[\ic{\bX^{obs} = \bx} - \prob{\bX = \bx} \right]\\
		&= \sum_x \sum_z \frac{\ic{\bX^{obs} = \bx}\ic{Z^{obs} = z+\delta}}{\prob{Z=z+\delta \mid \bX}} \left[Y - \expt{Y\mid \bX = \bx, Z=z+\delta}\right]\prob{Z=z\mid \bX}\\
		&\ \ + \sum_x \sum_z \expt{Y \mid \bX = \bx, Z = z+\delta} \ic{\bX^{obs}=\bx}\ic{Z^{obs} = z} \\
		&\ \ - \sum_x \sum_z \expt{Y \mid \bX = \bx, Z = z+\delta} \ic{\bX^{obs}=\bx}\prob{Z=z \mid \bX = \bx}\\
		&\ \ + \sum_x \sum_z \expt{Y \mid \bX = \bx, Z = z+\delta} \ic{\bX^{obs}=\bx}\prob{Z=z \mid \bX = \bx}\\
		&\ \ - \sum_x \sum_z\expt{Y \mid \bX = \bx, Z = z+\delta}\prob{Z = z\mid \bX = \bx}\prob{\bX = \bx}\\
		&= \left[Y - \expt{Y\mid \bX = \bX^{obs}, Z=Z^{obs}}\right]\frac{\prob{Z=Z^{obs}-\delta\mid \bX=\bX^{obs}}}{\prob{Z = Z^{obs}\mid \bX = \bX^{obs}}} \\
		&+ \expt{Y \mid \bX=\bX^{obs}, Z = Z^{obs}+\delta} - \expt{\mu(Z+\delta,\bX)}\\
		&\equiv \frac{\pi(Z-\delta\mid \bX)}{\pi(Z\mid \bX)}\left[Y - \mu(Z,\bX) \right] + \mu(Z+\delta,\bX) - \expt{\mu(Z+\delta,\bX)}
	\end{align}
	
	Going forward, we will denote the quantity above as:
	\begin{align}
		\xi(Y;\delta) \equiv \frac{\pi(Z-\delta\mid \bX)}{\pi(Z\mid \bX)}\left[Y - \mu(Z,\bX) \right] + \mu(Z+\delta,\bX)
	\end{align}
	
	and further take,
	\begin{align}
		\Xi(T; \delta, -\delta) \equiv \xi(T; \delta) - \xi(T; -\delta)
	\end{align}
	
	The equivalent procedure holds for $Z-\delta$ terms and for $A$.
	
	The above derivation assumes that $Z-\delta \in Supp(Z)$. This can be accomplished if $Z$ has infinite support or if we define our intervention functions in such a way to guarantee this. The approach we take throughout is to define interventions which leave the instrument at observed levels if the specified intervention would move them off the support. For example, if $Z \in [0, \infty)$, we might take the intervention to look like $Z \pm \delta \mathbb{I}\{ Z-\delta \geq 0 \}$
	
	This says we move $Z$ down by some amount $\delta$ unless such an intervention would create a negative $Z$ value. In this case, we leave $Z$ where it is and there is no effect on that unit.
	
	The form above shows the necessity of the strengthened positivity condition $\pi(Z-\delta\mid X)/\pi(Z\mid X) < \infty$ almost everywhere. 
	
	We repeat this analysis for the remaining regression functions (not shown) and then plugging in to (\ref{eq:chainrule}) gives:
	\begin{align}
		\varphi(z; \eta, \psi) &= \frac{g(\Xi(Y; \delta, -\delta) - f) - f(\Xi(A; \delta, - \delta) - g)}{g^2} \n
		&= \frac{\Xi(Y; \delta, -\delta) - \psi\Xi(A; \delta, - \delta)}{g} \n
		&=\frac{\Xi(Y; \delta, -\delta) - \psi\Xi(A; \delta, - \delta)}{\expt{\lambda(Z+\delta,
				\bX) - \lambda(Z-\delta,\bX)}} 
	\end{align}
	
	Now to estimate the parameter we take:
	\begin{align}
		\mathbb{P}_n (\varphi(z; \hat{\eta}, \hat{\psi})) &= 0 \n
		\Rightarrow \mathbb{P}_n \frac{\hat{\Xi}(Y; \delta, -\delta) - \hat{\psi} \hat{\Xi}(A; \delta, - \delta)}{\expt{\hat{\lambda}(Z+\delta,
				\bX) - \hat{\lambda}(Z-\delta,\bX)}}  &= 0 \n
		\Rightarrow \hat{\psi}_{IF} = \frac{\mathbb{P}_n(\hat{\Xi}(Y; \delta, - \delta)}{\mathbb{P}_n(\hat{\Xi}(A; \delta, -\delta))}&
	\end{align}
	
	We can now define the conditions under which our estimator is consistent, and give its asymptotic standard errors. Note that in this case, we are using $\delta$ and $-\delta$, below we give the proof in the bounded support case.
\end{proof}

\subsubsection{Derivation of the Efficient Influence Function with Bounded Support}\label{sec:BoundedSupport}
\begin{proof}
	We define our intervention where we shift the instrument level $Z$ up by some fixed amount $\delta$ as long as $Z+\delta \leq z_{max}$ and down by $\delta$ as long as $Z-\delta \geq z_{min}$. This gives a parameter:
	
	\begin{align}
		\psi^*_{shift} &= \frac{\expt{ \expt{Y \mid Z+\delta\ic{Z \leq z_{max} + \delta },X) - \expt{Y \mid Z-\delta\ic{Z \geq z_{min} - \delta },X)}}}}{\expt{ \expt{A \mid Z+\delta\ic{Z \leq z_{max} + \delta },X) - \expt{A \mid Z-\delta\ic{Z \geq z_{min} - \delta },X)}}}}\\
		&\equiv \frac{ \expt{\mu(Z+\delta\ic{Z \leq M_1},X) - \mu(Z-\delta\ic{Z \geq M_2},X)} }{ \expt{\lambda(Z+\delta\ic{Z \leq M_1},X) - \lambda(Z-\delta\ic{Z \geq M_2},X)} } \\
		&\equiv f/g
	\end{align}
	Starting with the $f$ term, we can break the intervention down depending on where $Z$ lies:
	
	\begin{align}
		f &= \mathbb{E}[ \left( \mu(Z+\delta) - \mu(Z-\delta) \right) \ic{M_2 \leq Z \leq M_1} \\ 
		&+  \left( \mu(Z+\delta) - \mu(Z) \right) \ic{Z \leq M_1,Z \leq M_2} \\ 
		&+  \left( \mu(Z) - \mu(Z-\delta) \right) \ic{Z \geq M_1,Z \geq M_2} ]\\
		\Rightarrow IF(f) &= \left[  \frac{\pi(Z-\delta)}{\pi(Z)} (Y - \mu(Z,X)) + \mu(Z+\delta,X) - \expt{\mu(Z+\delta,X)} \right. \\ \nonumber
		&\left.  -\frac{\pi(Z+\delta)}{\pi(Z)} (Y - \mu(Z,X)) - \mu(Z-\delta,X) + \expt{\mu(Z-\delta,X)} \right]\ic{M_2 \leq Z \leq M_1}\\
		&+ \left[  \frac{\pi(Z-\delta)}{\pi(Z)} (Y - \mu(Z,X)) + \mu(Z+\delta,X) - \expt{\mu(Z+\delta,X)} -  (Y - \expt{\mu(Z)})  \right]\ic{Z \leq M_1, Z\leq M_2}\\
		&+ \left[  (Y - \expt{\mu(Z,X)}) - \frac{\pi(Z+\delta)}{\pi(Z)} (Y - \mu(Z,X)) - \mu(Z-\delta,X) + \expt{\mu(Z-\delta,X)} \right]\ic{Z \geq M_1,Z \geq M_2}\\
		&= \left[ \xi(Y; \delta) - \xi(Y; -\delta)- \expt{\mu(Z+\delta) - \mu(Z-\delta)} \right] \ic{Z \leq M_1, Z \geq M_2}\\
		&+ \left[ \xi(Y; \delta) - \xi(Y;0)- \expt{\mu(Z+\delta) - \mu(Z)} \right] \ic{Z \leq M_1,Z \leq M_2}  \\
		&+ \left[ \xi(Y;0)- \xi(Y; -\delta)- \expt{ \mu(Z+\delta,X) - \mu(Z,X)} \right]\ic{Z \geq M_1,Z \geq M_2} \\
		&= \xi(Y; \delta\ic{Z \leq M_1}) - \xi(Y; \delta\ic{Z\geq M_2}) -f\\
		&\equiv \Xi(Y; \delta\ic{Z \leq M_1}, -\delta\ic{Z\geq M_2} ) - f
	\end{align}
	
	Note that in the single shift case, this reduces very considerably as follows:
	\begin{align}
		\xi(Y; \delta \ic{Z \leq M_1}) - \xi(Y;0) = \xi(Y; \delta)\ic{Z \leq M_1}
	\end{align}
	
	The same holds for the $g$ term, replacing $Y$ with $A$. Now we can find the IF of the parameter itself:
	
	\begin{align}
		IF(\psi) &= \left[ ( \Xi(Y; \delta\ic{Z \leq M_1}, -\delta\ic{Z\geq M_2} ) - f)g - ( \Xi(A; \delta\ic{Z \leq M_1}, -\delta\ic{Z\geq M_2} ) - g)f\right] / g^2 \\
		&= \left[  \Xi(Y; \delta\ic{Z \leq M_1}, -\delta\ic{Z\geq M_2} )  - \psi  \Xi(A; \delta\ic{Z \leq M_1}, -\delta\ic{Z\geq M_2} ) \right] / g
	\end{align}
	
	This gives the IF-based estimator of:
	
	\begin{align}
		\hat{\psi}^*_{IF} &= \frac{ \mathbb{P}_n \left[\ \Xi(Y; \delta\ic{Z \leq M_1}, -\delta\ic{Z\geq M_2} ) \right] } { \mathbb{P}_n \left[  \Xi(A; \delta\ic{Z \leq M_1}, -\delta\ic{Z\geq M_2} )\right]}
	\end{align}

\end{proof}

\subsubsection{Proof of Convergence (Theorem \ref{thm:DoubleDR})}
\begin{proof}\label{pf:SingleVar}
	In an abuse of notation, rewrite the estimator as:
	\begin{align}
		\hat{\psi}_{IF} &= \frac{P_n f(Y;\hat{\eta})}{P_n g(A;\hat{\eta})} 
	\end{align}
	
	And likewise the true parameter is written as:
	\begin{align}
		\psi_{shift} &= \frac{P f(Y;\eta)}{P g(A;\eta)} 
	\end{align}
	Now we take:
	\begin{align}
		\hat{\psi}_{IF} - \psi_{shift} &= \frac{P_n f(Y;\hat{\eta})}{P_n g(A;\hat{\eta})}  - \frac{P f(Y;\eta)}{P g(A;\eta)} \\ \nonumber
		&= \frac{P g(A;\eta)P_n f(Y;\hat{\eta}) - P f(Y;\eta)P_n g(A;\hat{\eta})}{P_n g(A;\hat{\eta})P g(A;\eta)} \\ \nonumber
		&= \frac{P g(A;\eta) \left\{ P_nf(Y;\hat{\eta}) - P f(Y;\eta)  \right\} -Pf(Y;\eta)\left\{ P_n g(A;\hat{\eta}) -P g(A;\eta) \right\} }{P_n g(A;\hat{\eta})P g(A;\eta)} \\ \nonumber
		&= \frac{1}{P_n g(A;\hat{\eta})} \left[P_n f(Y;\hat{\eta}) -Pf(Y;\eta)) - \psi_{shift} (P_n g(A;\hat{\eta}) - P  g(A;\hat{\eta}))\right]
	\end{align}
	We can treat the first ratio as a constant under the causal assumptions, and break down the other components into smaller pieces, for example:
	\begin{align}
		P_n f(Y;\hat{\eta}) - P f(Y;\eta) &= (P_n - P)( f(Y;\hat{\eta}) -  f(Y;\eta)) + (P_n - P)  f(Y;\eta) - P( f(Y;\hat{\eta}) - f(Y;\eta) )
	\end{align}
	The first term on the right is $o_p(1/\sqrt{n})$ by van der Vaart (2000) as long as we assume Donsker or by estimating the nuisance parameters on a separate sample. The second term is normal by the CLT. The third term needs to be bounded by iterated expectation as follows. First we take the expectation over $X$ and $Z$:
	\begin{align}
		P( f(Y;\hat{\eta}) - f(Y;\eta) ) &= \mathbb{E}\left( \frac{\hat{\pi}(Z-\delta \mid X)}{\hat{\pi}(Z \mid X)} (\mu(Z,X) - \hat{\mu}(Z,X)) + \hat{\mu}( Z+\delta,X) \right) \n
		&+ \mathbb{E}\left( \frac{\hat{\pi}(Z+\delta \mid X)}{\hat{\pi}(Z \mid X)} (\mu(Z,X) - \hat{\mu}(Z,X)) + \hat{\mu}( Z-\delta,X) \right) \n
		&  - \mathbb{E}\left\{\mu( Z+\delta,X) - \mu(Z-\delta,X)\right\} 
	\end{align}
	
	In this, we assume that either $Z$ has infinite support of the intervention functions have been properly specified such that the terms where $Z\pm \delta$ is outside of the support of $Z$ reduce to 0. Looking at the first line:
	\begin{align}
		\mathbb{E}&\left( \frac{\hat{\pi}(Z-\delta \mid X)}{\hat{\pi}(Z\mid X)} ( \mu(Z,X) - \hat{\mu}(Z,X))\right)  + \mathbb{E}\left\{ \hat{\mu}( Z+\delta,X) -  \mu(Z+\delta,X)  \right\}
	\end{align}
	Integrate over $z$
	\begin{align*}
		&= \mathbb{E}\left(  \int \frac{\hat{\pi}(Z-\delta \mid X)}{\hat{\pi}(Z\mid X)} ( \mu(Z,X) - \hat{\mu}(Z,X))\pi(Z\mid X) dz\right. \n
		&\left.+ \int (\hat{\mu}( Z+\delta,X) -  \mu(Z+\delta,X)  )\pi(Z\mid X)dz  \right) \n
	\end{align*}
	Change of variables:
	\begin{align*}
		&= \mathbb{E}\left(  \int \frac{\hat{\pi}(Z\mid X)}{\hat{\pi}( Z+\delta \mid X)} ( \mu(Z+\delta,X) - \hat{\mu}( Z+\delta,X))\pi( Z+\delta \mid X) dz +\right. \n
		& \left.\int (\hat{\mu}( Z+\delta,X) -  \mu(Z+\delta,X)  )\pi(Z\mid X)dz  \right) \n
		&= \mathbb{E}\left(  \int \left( \frac{\hat{\pi}(Z\mid X)/\hat{\pi}( Z+\delta \mid X)}{\pi(Z\mid X)/\pi( Z+\delta \mid X)} ( \mu(Z+\delta,X) - \hat{\mu}( Z+\delta,X))\pi(Z\mid X)\right.\right. \n
		&\left. \left. + (\hat{\mu}( Z+\delta,X) -  \mu(Z+\delta,X)  )\pi(Z\mid X) \right) dz  \right) \n
		&= \mathbb{E}\left(  \int \frac{\hat{\pi}(Z\mid X)/\hat{\pi}( Z+\delta \mid X) - \pi(Z\mid X)/\pi( Z+\delta \mid X)}{\pi(Z\mid X)/\pi( Z+\delta \mid X)}\right. \n
		&\left. ( \mu(Z+\delta,X) - \hat{\mu}( Z+\delta,X))\pi(Z\mid X) dz   \right) \n
		&= \mathbb{E}\left( \frac{\hat{\pi}(Z\mid X)/\hat{\pi}( Z+\delta \mid X) - \pi(Z\mid X)/\pi( Z+\delta \mid X)}{\pi(Z\mid X)/\pi( Z+\delta \mid X)}( \mu(Z+\delta,X) - \hat{\mu}( Z+\delta,X))  \right)  \n    
	\end{align*}
	
	Note that we can choose how to do the change of variables step. We could equally say:
	\begin{align}
		&\mathbb{E}\left(  \int \frac{\hat{\pi}(Z-\delta \mid X)}{\hat{\pi}(Z\mid X)} ( \mu(Z,X) - \hat{\mu}(Z,X))\pi(Z\mid X) dz \right. \n
		&\left. + \int (\hat{\mu}( Z+\delta,X) -  \mu(Z+\delta,X)  )\pi(Z\mid X)dz  \right) \n
		&= \mathbb{E}\left(  \int \frac{\hat{\pi}(Z -\delta \mid X)}{\hat{\pi}( Z \mid X)} ( \mu(Z,X) - \hat{\mu}( Z,X))\pi( Z \mid X) dz +\right. \n
		& \left.\int (\hat{\mu}( Z,X) -  \mu(Z,X)  )\pi(Z-\delta \mid X)dz  \right) \n
	\end{align}
	This will give a final result:
	\begin{align}
		\mathbb{E}\left( \frac{\hat{\pi}(Z -\delta \mid X)/\hat{\pi}( Z \mid X) - \pi(Z -\delta \mid X)/\pi( Z \mid X)}{\pi(Z -\delta \mid X)/\pi( Z \mid X)}( \mu(Z,X) - \hat{\mu}( Z,X))  \right) 
	\end{align}
	
	The first formulation requires $\pi(Z\mid X)/\pi( Z+\delta \mid X)$ and its empirical estimate to be bounded away from zero and the second requires that same condition hold for $\pi(Z - \delta \mid X)/\pi( Z \mid X)$. Assuming these, we can bound this quantity using Cauchy Schwartz:
	
	\begin{align}
		\lesssim ||\frac{\hat{\pi}(Z\mid X)}{\hat{\pi}( Z+\delta \mid X)} - \frac{\pi(Z\mid X)}{\pi( Z+\delta \mid X)}|| \cdot || \mu(Z+\delta,X) - \hat{\mu}( Z+\delta,X)|| 
	\end{align} 
	Or, equivalently,
	\begin{align}
		\lesssim ||\frac{\hat{\pi}(Z-\delta \mid X)}{\hat{\pi}( Z\mid X)} - \frac{\pi(Z-\delta \mid X)}{\pi( Z \mid X)}|| \cdot || \mu(Z,X) - \hat{\mu}( Z,X)|| 
	\end{align} 
	
	We can repeat this process with the $Z-\delta$ terms, and the same proof holds for the $\lambda$ terms, with the same requirement of positivity in the propensity score ratios.
	
	To achieve consistent estimation of $\psi$, we therefore need 
	
	\begin{align}
		&||\frac{\hat{\pi}(Z - \delta \mid X)}{\hat{\pi}(Z \mid X)} - \frac{\pi(Z - \delta \mid X)}{\pi(Z \mid X)}|| \cdot (||\mu(Z+\delta,X) - \hat{\mu}(Z+\delta,X)|| + ||\lambda(Z+\delta, X) - \hat{\lambda}(Z+\delta, X)||) \n
		&+ ||\frac{\hat{\pi}(Z+\delta \mid X)}{\hat{\pi}(Z\mid X)} - \frac{\pi(Z+\delta \mid X)}{\pi(Z\mid X)}|| \cdot (||\mu(Z-\delta,X) - \hat{\mu}(Z-\delta,X)|| + ||\lambda(Z-\delta, X) - \hat{\lambda}(Z-\delta, X)||) \n
		&= o_p(1/\sqrt{n})
	\end{align} 
	
	This is satisfied if all terms are estimated at faster than $n^{1/4}$ or if the propensity score ratios (or both sets of regression functions) are estimated at $n^{1/2}$. 
	
\end{proof}

\subsubsection{Proof of uniform convergence (Theorem \ref{thm:Uniform})} 
\% Proof for uniform confidence intervals

\begin{proof}
	We closely follow the proof in \citet{kennedy2019nonparametric}, refer there for details. The goal is to provide a bound for the quantity
	
	\begin{align}
		\sup_{\delta \in \mathcal{D}} \frac{\sqrt{n}(\hat{\psi}_{IF}(\delta) - \psi_{shift}(\delta))}{\hat{\sigma}(\delta)} - \sqrt{n}(\mathbb{P}_n - \mathbb{P})\left( \frac{\varphi(z; \eta, \delta)}{\sigma(\delta)}  \right)
	\end{align}
	
	Let the norm $||f||_{\mathcal{D}}$ signify $sup_{\delta \in \mathcal{D}}$. First, if $\varphi$ is Lipschitz and $||\frac{\hat{\sigma}}{\sigma} -1 ||_{\mathcal{D}} = o_p(1)$, we can rewrite the quantity above as
	
	\begin{align}
		|| \frac{\sqrt{n}(\hat{\psi}_{IF}(\delta) - \psi_{shift}(\delta))}{\sigma(\delta)} - \sqrt{n}(\mathbb{P}_n - \mathbb{P})\left( \frac{\varphi(z; \eta, \delta)}{\sigma(\delta)}  \right) ||_{\mathcal{D}} + o_p(1)
	\end{align}
	
	Now we split the above quantity into one piece which accounts for estimation error, and one which captures the error in nuisance function estimation. We do this by splitting the data over $K$ folds. 
	
	\begin{align}
		\frac{\sqrt{n}(\hat{\psi}_{IF}(\delta) - \psi_{shift}(\delta))}{\sigma(\delta)} &- \sqrt{n}(\mathbb{P}_n - \mathbb{P})\left( \frac{\varphi(z; \eta, \delta)}{\sigma(\delta)}  \right) \n &= \frac{\sqrt{n}}{K\sigma(\delta)} \sum_{k=1}^{K} \sqrt{\frac{K}{n}} (\mathbb{P}_n^k - \mathbb{P}) \{\varphi(z; \hat{\eta}_{-k} - \varphi(z; \eta, \delta))  \} - \mathbb{P}\{ \varphi(z; \hat{\eta}_{-k}, \delta) - \varphi(z; \eta,\delta)  \} \n
		&= B_{n,1}(\delta) + B_{n,2}(\delta)
	\end{align}
	
	The proof given in \cite{kennedy2019nonparametric} to bound $||B_{n,1}(\delta)||_{\mathcal{D}}$ is completely general, so we do not give it here. It shows 
	
	\begin{align}
		||B_{n,1}(\delta)||_{\mathcal{D}} &= o_p(1)
	\end{align}
	
	The second task is to bound $||B_{n,2}(\delta)||_{\mathcal{D}}$. We have already shown,
	
	\begin{align}\label{eq:DRresult}
		\mathbb{P}\{ \Xi(Y; \delta, -\delta) - \hat{\Xi}(Y; \delta, -\delta) \} &\lesssim ||\frac{\hat{\pi}(Z-\delta \mid X)}{\hat{\pi}(Z\mid X)} - \frac{\pi(Z-\delta)}{\pi(Z \mid X)}|| \cdot ||\mu(Z+\delta,X) - \hat{\mu}(Z+\delta,X)||\\
		&+ ||\frac{\hat{\pi}(Z + \delta \mid X)}{\hat{\pi}(Z\mid X)} - \frac{\pi(Z+\delta)}{\pi(Z \mid X)}|| \cdot ||\mu(Z-\delta,X) - \hat{\mu}(Z-\delta,X)||
	\end{align}
	
	in $L_2(P)$ norm. The same holds replacing $Y$ with $A$ and $\mu$ terms with $\lambda$ terms. 
	
	Recall that our influence function $\varphi$ is given by,
	
	\begin{align}
		\varphi &= \frac{\Xi(Y; \delta, -\delta) - \psi_{shift}\Xi(A; \delta, -\delta)}{\mathbb{E}(\lambda(Z+\delta, X) - \lambda(Z-\delta,X))}
	\end{align}
	
	Thus,
	
	\begin{align}
		\mathbb{P}\{ \varphi(z; \hat{\eta}, \delta) - \varphi(z; \eta,\delta)  \} &= \mathbb{P}\{ \Xi(Y;\delta, -\delta) - \hat{\Xi}(Y; \delta, -\delta) - \psi_{shift}\Xi(A; \delta, -\delta) + \hat{\psi}_{IF}\hat{\Xi}(Y; \delta, -\delta)  \} \n
		&\lesssim ||\frac{\hat{\pi}(Z - \delta \mid X)}{\hat{\pi}(Z \mid X)} - \frac{\pi(Z-\delta \mid X)}{\pi(Z\mid X)}|| \n
		&\cdot \left(  ||\mu(Z+\delta,X) - \hat{\mu}(Z+\delta,X)|| +||\lambda(Z+\delta,X) - \hat{\lambda}(Z+\delta,X)||  \right)\\
		&+ ||\frac{\hat{\pi}(Z + \delta \mid X)}{\hat{\pi}(Z \mid X)} - \frac{\pi(Z+\delta \mid X)}{\pi(Z\mid X)}|| \n
		&\cdot \left(  ||\mu(Z-\delta,X) - \hat{\mu}(Z-\delta,X)|| +||\lambda(Z-\delta,X) - \hat{\lambda}(Z-\delta,X)||  \right)
	\end{align}
	
	where the $\psi_{shift}$ terms and the denominator are constant, and therefore drop out. If we make the assumption that the supremum of the quantity above across all $\delta$ in some $\mathcal{D}$ is $o_p(1/\sqrt{n})$, the bound for $B_{n,2}(\delta)$ is complete. Thus the entire quantity is bounded uniformly across $\delta$.
	
\end{proof}
\end{document}